\documentclass[a4paper,11pt]{article}
\usepackage{jinstpub} 
\usepackage{lineno}
\usepackage[amssymb]{SIunits}
\usepackage[T1]{fontenc}
\usepackage{subcaption}

\title{\boldmath Timing resolution performance of Timepix4 bump-bonded assemblies}

\author[a,b,1]{R.~Bolzonella \note{Corresponding author.}}
\author[c]{J.~Alozy}
\author[c]{R.~Ballabriga}
\author[d]{M.~van~Beuzekom}
\author[a]{N.~V.~Biesuz}
\author[c]{M.~Campbell}
\author[a]{P.~Cardarelli}
\author[a,b]{V.~Cavallini}
\author[c]{V.~Coco}
\author[a]{A.~Cotta~Ramusino}
\author[a,b]{M.~Fiorini}
\author[d]{V.~Gromov}
\author[a,b]{M.~Guarise}
\author[c]{X.~Llopart Cudie}
\author[a,b]{S.~Okamura}
\author[a,b]{G.~Romolini}
\author[a,b]{A.~Saputi}
\author[d]{A.~Vitkovskiy}

\affiliation[a]{INFN Ferrara, Via Saragat 1, 44122 Ferrara, Italy}
\affiliation[b]{University of Ferrara, Via Saragat 1, 44122 Ferrara, Italy}
\affiliation[c]{CERN, 1211 Geneva 23, Switzerland}
\affiliation[d]{Nikhef, Science Park 105, 1098 XG Amsterdam, the Netherlands}

\emailAdd{rbolzonella@fe.infn.it}

\abstract{
The timing performance of the Timepix4 application-specific integrated circuit (ASIC) bump-bonded to a $100\;\micro\meter$ thick n-on-p silicon sensor is presented. A picosecond pulsed infrared laser was used to generate electron-hole pairs in the silicon bulk in a repeatable fashion, controlling the amount, position and time of the stimulated charge signal. The timing resolution for a single pixel has been measured to $107\;\pico\second$ r.m.s. for laser-stimulated signals in the silicon sensor bulk. Considering multi-pixel clusters, the measured timing resolution reached $33\;\pico\second$ r.m.s. exploiting oversampling of the timing information over several pixels.
}


\begin{document}
\maketitle
\flushbottom

\section{Introduction}\label{sec:introduction}

Timepix4 is the latest generation application-specific integrated circuit (ASIC) of the Timepix family~\cite{bib:Llopart_2022,bib:Ballabriga:2018lzk}, mainly targeted for single particle detection in hybrid pixel detectors~\cite{bib:Llopart_2022}.

The Timepix4 will be used in a novel device for the detection of single visible photons combining excellent timing and spatial resolution with high rate capabilities and large active area, currently under development within the 4DPHOTON project~\cite{bib:Fiorini_2018,Alozy:2021kqn}: it is based on a vacuum tube, with transmission photocathode, a microchannel plate (MCP) for electron multiplication and the Timepix4 ASIC used as pixelated anode.
This kind of approach, exploiting a bare ASIC used inside a vacuum tube, has been already proved on other works \cite{bib:Vallerga:2008bbo, bib:Vallerga:2014wwa}.
A detector with such performance would have a fundamental role in improving state-of-the-art imaging technology, with applications in many fields such as high-energy physics, life sciences or quantum physics to quote a few examples.

A detailed characterization of the Timepix4 ASIC, especially in terms of timing resolution, is necessary to correctly assess the performance of the MCP-based hybrid photon detector, which is currently under construction. This paper presents the timing characterization of the available assembly of the Timepix4 bump-bonded to a $100\;\micro\meter$ thick n-on-p Si detector illuminated by a picosecond infrared laser.

\section{Timepix4 ASIC}\label{sec:Timepix4}

Timepix4 is a CMOS ASIC in $65\;\nano\meter$ developed by the Medipix4 Collaboration for the readout of pixel detectors, which is able to reach high spatial and timing resolutions together with high rate capabilities.

\begin{figure}[htbp]
\centering
\includegraphics[width=.9\textwidth]{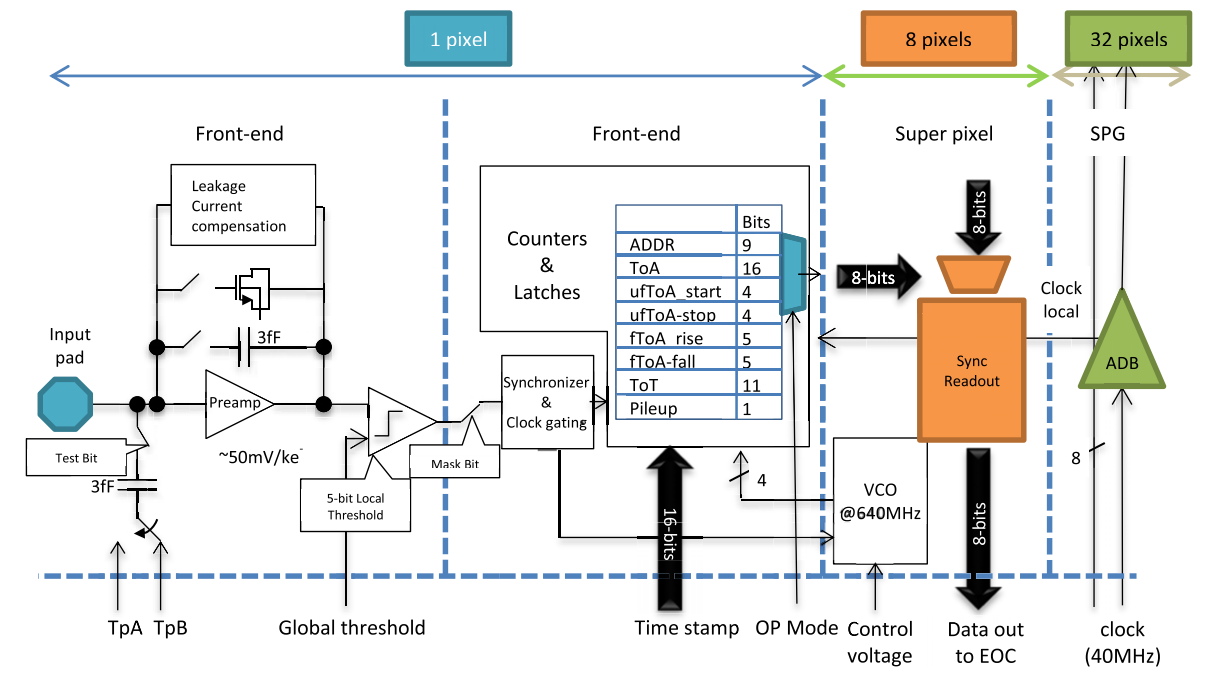}
\caption{Simplified block diagram of the pixel front-end electronics and the pixel, SP and SPG structure. \protect\cite{bib:Llopart_2022}\label{fig:block_diagram_structure}}
\end{figure}

The Timepix4 consists of a $512\times448$ array with square pixels with a $55\;\micro\meter$ pitch, for a total active area of about $7\;\centi\meter^2$. 
The Timepix4 front-end can be configured to be sensitive either to electrons or holes. 
Pixel electronics amplifies and discriminates the signals in each pixel, as shown in Figure~\ref{fig:block_diagram_structure}. Signals which exceed the threshold are processed by the digital front-end. A 5-bit Digital-to-Analog Converter (DAC) per pixel allows to equalize the threshold value.
\newline A Time to Digital Converter (TDC) in each pixel uses a Voltage-Controlled Oscillator (VCO) to measure the Time-of-Arrival (ToA) and Time-over-Threshold (ToT) for each signal above threshold. The TDC nominal bin size is $195\;\pico\second$ that corresponds to a $56\;\pico\second$ r.m.s. resolution.
\newline The pixels are organized in groups of 8 (2 columns of 4 pixels), called SuperPixels (SP), sharing the same TDC and the same VCO. 4 SuperPixels in one double column form a SuperPixel Group (SPG) and 16 SPGs form a full double column.

\medskip
Timepix4 can be operated in data-driven or frame-based mode. The measurements presented in this paper are based on the data-driven mode: each single pixel hit produces a 64-bit data word, 64b/66b encoded, which is transmitted off-chip by 16 high speed programmable links running at a maximum $10\;\giga\textrm{b}\per\second$ data bandwidth. Therefore, the Timepix4 ASIC can reach a maximum bandwidth of $160\;\giga\textrm{b}\per\second$ corresponding to a maximum hit rate of $\sim2.5\;\giga\textrm{Hit}\per\second$. 

The Timepix4 ensures also a low equivalent noise charge ($\sim80\;\textrm{e}^-$) per pixel, which, combined with a threshold dispersion after equalization of $\sim40\;\textrm{e}^-$, allows to set a low threshold ($\sim 800 \textrm{e}^-$) to the whole ASIC masking just few noisy pixels \cite{bib:Llopart_2022}.

External digital signals can be sent to the Timepix4 through two single-ended inputs (called "digital pixels") in the top and the bottom peripheries. At one periphery one external input is routed to the even columns and the other to the odd columns. One SP (at the end of the column) is chosen to receive each of the external input signals. In the chosen SP signals for all 8 front-ends are ignored. This permits up to 4 external inputs to be included in the output data stream at the expense of losing 8 pixel inputs per external input.

\subsection{Time measurement}\label{subsec:Time_measurement}

\begin{figure}[htbp]
\centering
\subfloat[]
{
    \centering
    \includegraphics[width=.48\textwidth]{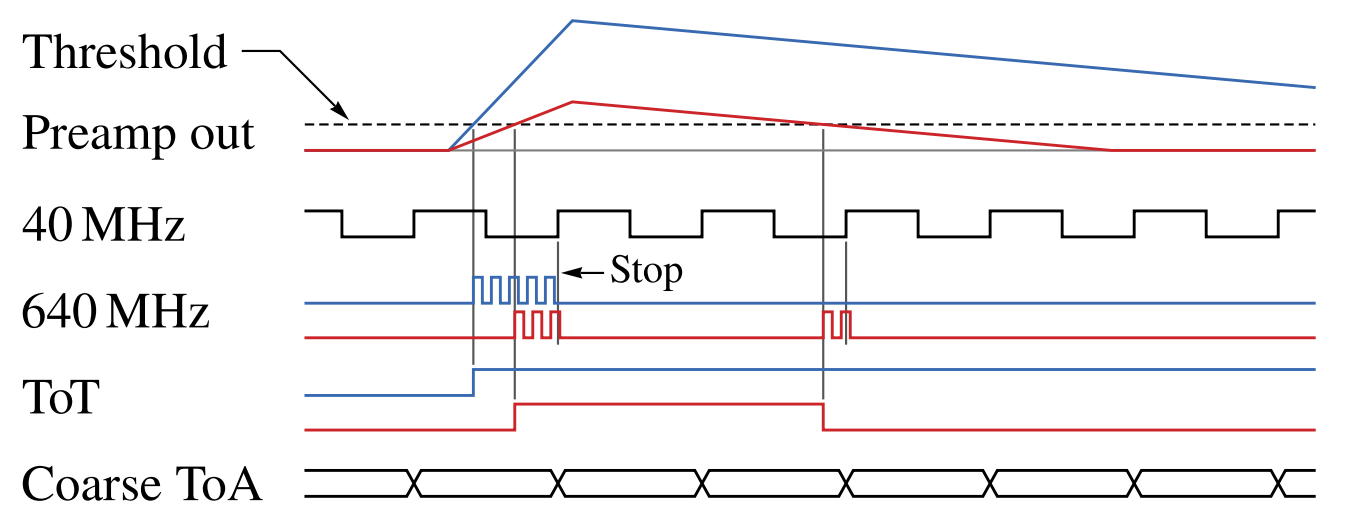}
    \label{fig:ToA_operation}
}
\hfill
\subfloat[]
{
    \centering
    \includegraphics[width=.48\textwidth]{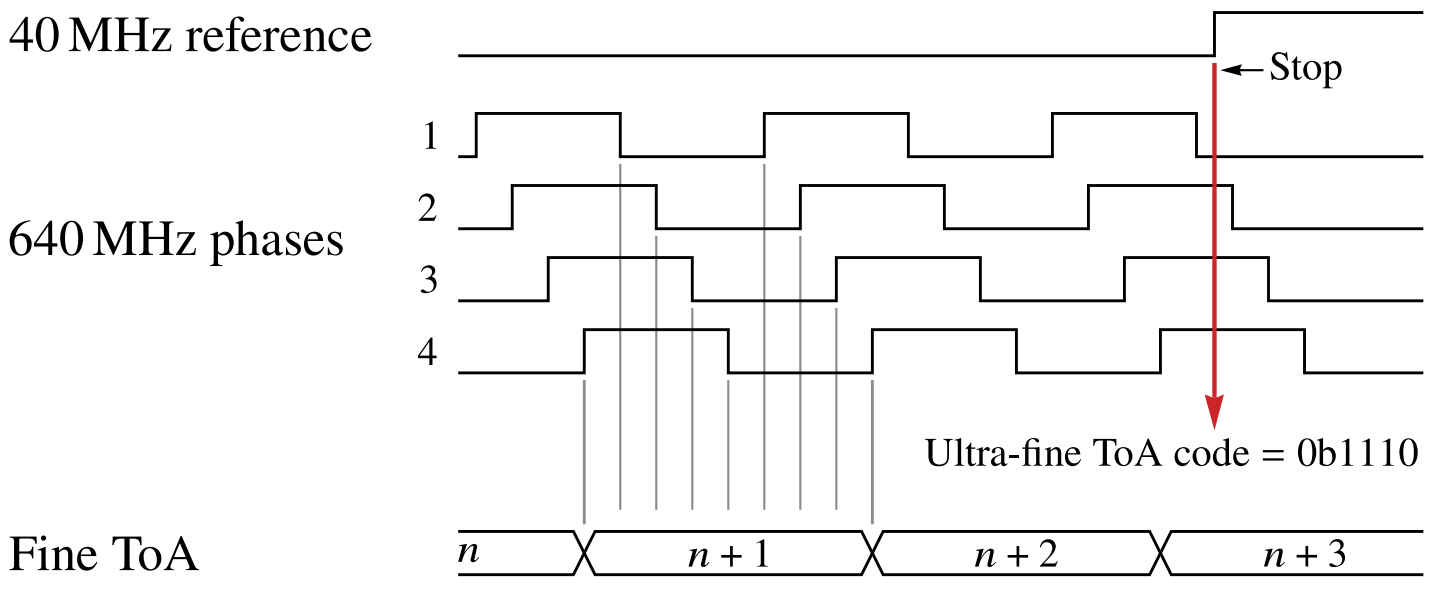}
    \label{fig:UfToA_operation}
}
\caption{\textbf{(a)} Operation of the ToA and fToA measurement. \textbf{(b)} Copies of $640\;\mega\hertz$ clock shifted returning ufToA measurement \protect\cite{bib:Heijhoff_2022}. \label{fig:ToA_clocks_operation}}
\end{figure}

The time-to-digital conversion in the Timepix4 ASIC happens in three steps \cite{bib:Heijhoff_2022}. The first one consists of the determination of the $40\;\mega\hertz$ clock cycle in which the preamplifier output signal rises above threshold. When this happens, a clock, generated by the SuperPixel VCO, is enabled. This clock has a nominal frequency of $640\;\mega\hertz$, providing a fine-Time-of-Arrival (fToA), as shown in Figure~\ref{fig:ToA_operation}. This clock is active until the following $40\;\mega\hertz$ clock rising edge, so the number of $640\;\mega\hertz$ clock cycles provides a finer measurement of the ToA, improving the bin width from $25\;\nano\second$ to $1.56\;\nano\second$.
Moreover, when the VCO is activated, 4 internal phases of the oscillator can be latched to further divide the $1.56\;\nano\second$ bins into 8 bins with nominal width of $195\;\pico\second$ and providing an ultra-fine-Time-of-Arrival (ufToA), as shown in Figure~\ref{fig:UfToA_operation}.

\section{Experimental setup}\label{sec:setup}

The measurements presented in the paper consist of sending simultaneous signals to different pixels and measuring the time-difference between them to estimate the timing resolution.
In order to do that, a pulse generator (Active Technologies PG-1072) with a measured interchannel jitter of $7\;\pico\second$ r.m.s. has been used to produce two synchronized pulses with a period of $5\;\milli\second$.

\medskip

\begin{figure}[htbp]
\centering
\includegraphics[width=.95\textwidth]{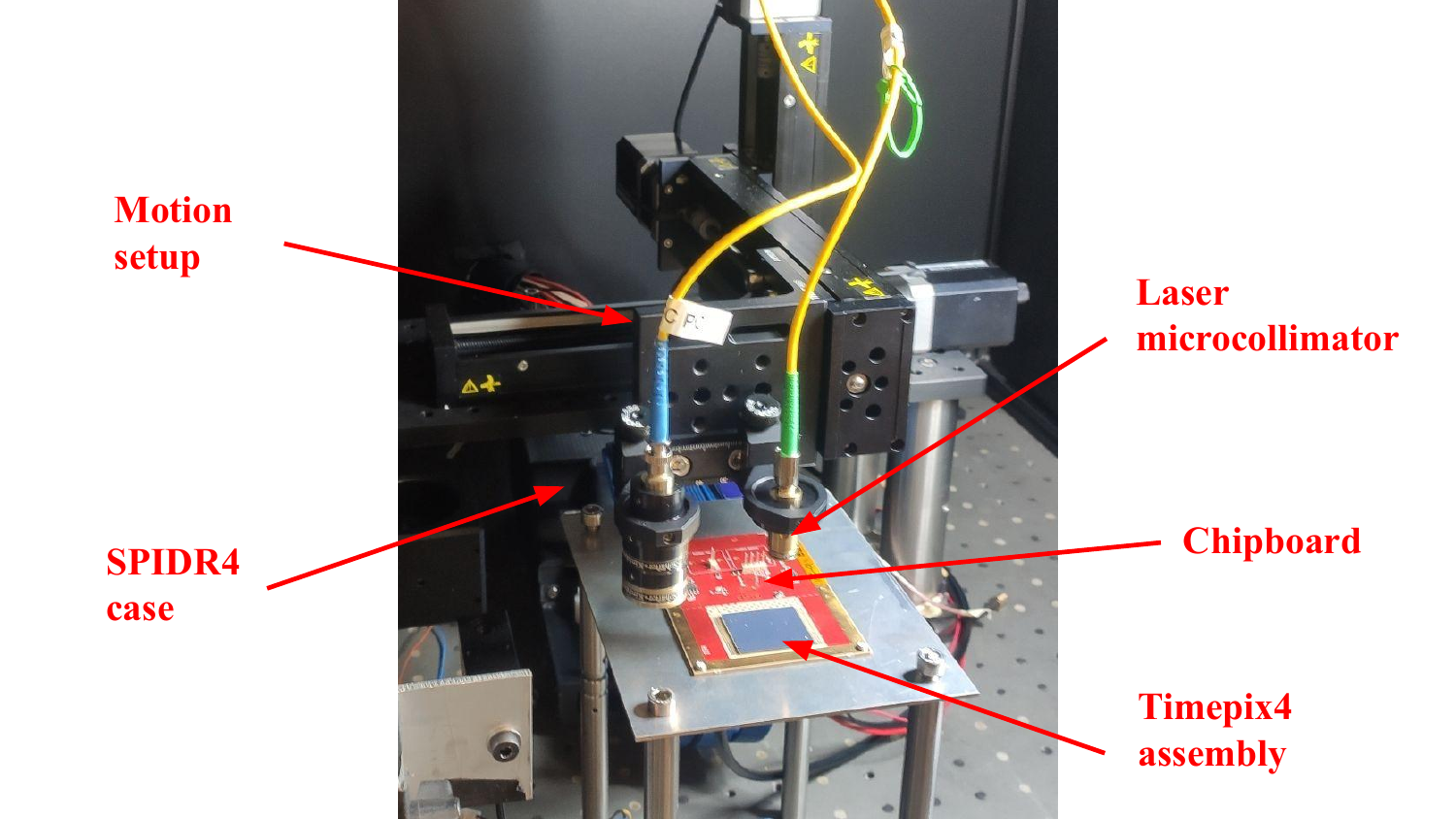}
\caption{Laser setup used for the described measurements.\label{fig:setup}}
\end{figure}

The first pulse is sent directly to one of the Timepix4 digital pixels. 
The second pulse is used to trigger a pulsed diode laser (PicoQuant PDL-800 B driver with LDH-P-1060 laser head), which produces an infrared laser pulse that is then attenuated and guided to the Timepix4 through an optical fiber patch cord. This reaches a microcollimator, which is positioned using through linear translation stages with micrometric precision. The overall setup is shown in Figure~\ref{fig:setup}.

\subsection{DAQ system: Timepix4 assembly and SPIDR4 control board}
\begin{figure}[htbp]
\centering
\subfloat[]
{
    \centering
    \includegraphics[width=.5\textwidth]{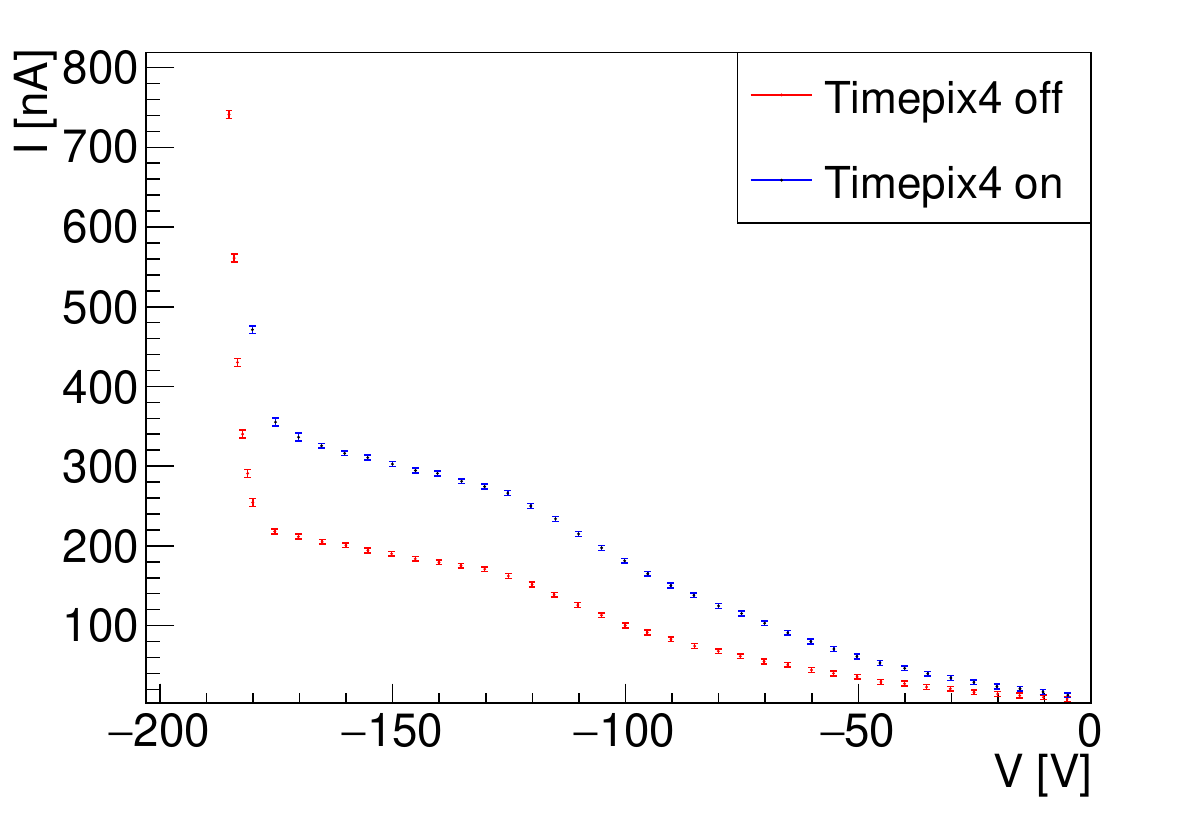}
    \label{fig:IV_curve}
}
\hfill
\subfloat[]
{
    \centering
    \includegraphics[width=.45\textwidth]{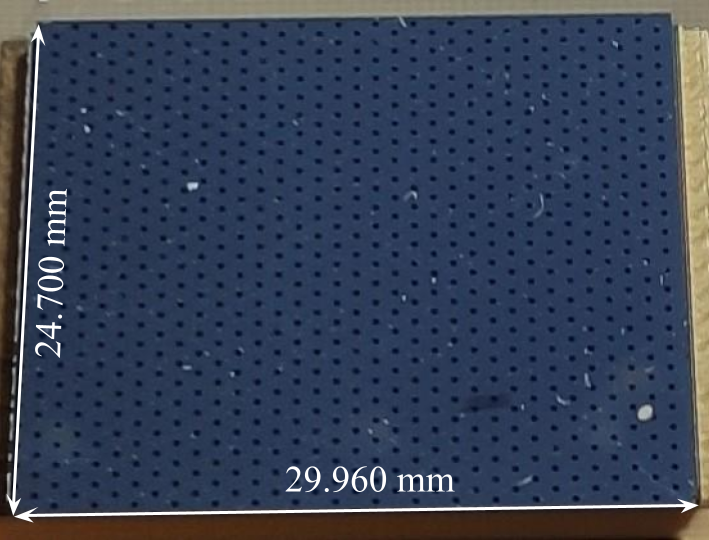}
    \label{fig:metalization}
}
\caption{\textbf{(a}) I-V dependency of the Si sensor bonded to the Timepix4. \textbf{(b)} Holes pattern in the metallization coated on the Si sensor. The holes have a diameter of $300\;\micro\meter$ and they are disposed in a triangular mesh spaced by $\;\milli\meter$}
\end{figure}

The Timepix4 ASIC used in this measurements is bump-bonded to a $100\;\micro\meter$-thick n-on-p silicon detector, and glued and wire-bonded onto a carrier board developed by Nikhef. The silicon sensor has a breakdown voltage of ~$-180\;\volt$, as shown in Figure \ref{fig:IV_curve}, and it is biased at $-150\;\volt$ for the measurements presented here. The current flowing in the sensor is higher if the Timepix4 is switched on, due to a higher temperature, but the breakdown voltage is not affected by that. The sensor has a back-side metallization with a pattern of circular openings with 300 micron diameter (visible in Figure \ref{fig:metalization}) which allow the laser light to enter the silicon bulk. The silicon attenuation coefficient for a wavelength $1060\;\nano\meter$ is about 900 microns, which ensures an almost uniform generation of electron-hole pairs through the silicon bulk thickness.

Measurements at very low energies are not required for the purpose of this paper. Low energy measurements are strongly affected by the time-walk effect, while they only give a small improvement to the overall resolution of the cluster. Hence these noise-pixels are discarded by setting a threshold corresponding to several sigmas of the Equivalent Noise Charge (ENC). 
In the measurements described in this paper, the Timepix4 is operated in electron-collecting mode with a threshold set to $1000\;\textrm{e}^-$ after pixel equalization. 
The DACs configuration set, reported in Table \ref{tab:DAC_settings}, leads to a pixel current of $7.67\;\micro\ampere$. By increasing the power per pixel the timing resolution would improve at low input charges, where the resolution is dominated by the front-end contribution \cite{bib:Heijhoff_2022}.

\begin{table}[!ht]
\begin{subtable}[t]{0.45\textwidth}
\centering
\begin{tabular}[t]{|c|c|c|}
\hline
\textbf{DAC name} & \textbf{DAC value set} & \textbf{DAC gain set}\\ \hline
VcascPreamp & 170 & 5\\ \hline
VFBK & 128 & 2\\ \hline
VCascDisc & 130 & 5\\ \hline
VControlVCO & 0 & 2\\ \hline
VThreshold & 8192 & 5\\ \hline
VTPCoarse & 128 & 2\\ \hline
VTPFine & 512 & 5\\ \hline
\end{tabular}
\caption{Bias voltage DACs.}
\label{subtab:V_DACs_settings}
\end{subtable}
\hspace{\fill}
\begin{subtable}[t]{0.45\textwidth}
\centering
\begin{tabular}[t]{|c|c|}
\hline
\textbf{DAC name} & \textbf{DAC value set}\\ \hline
VBiasPreamp & 100\\ \hline
VbiasDiscTailNMOS & 60\\ \hline
VBiasIkrum & 5\\ \hline
VBiasLevelShift & 90\\ \hline
VbiasDiscPMOS & 40\\ \hline
VBiasDiscTRAFF & 50\\ \hline
VBiasADC & 128\\ \hline
VBiasDAC & 50\\ \hline
\end{tabular}
\caption{Bias current DACs.}
\label{subtab:I_DACs_settings}
\end{subtable}
\caption{DAC settings of the Timepix4 used in all the measurements presented in this paper.}
\label{tab:DAC_settings}
\end{table}

The Timepix4 carrier board is connected through a flat cable to a SPIDR4 control board, developed by Nikhef, used both to configure the ASIC and to transfer the output data to a data acquisition PC. During the measurements, two high-speed links were used (one link per half matrix) both running at $2.56\;\giga\textrm{b}\per\second$.

\subsection{Laser setup and configuration}\label{subsec:laser}

The infrared laser pulse has a full-width-half-maximum time duration of about $36\;\pico\second$, and a jitter with respect to the laser trigger and synch-output signals lower than $20\;\pico\second$ r.m.s. from datasheet.
The amount of light per single pulse is tuned using a variable fiber optics attenuator (VOA Electronics V1000F). The light is then collimated and focused in order to illuminate the smallest possible region onto the sensor back-side. Precise positioning and focusing of the laser spot in the center of the metallization openings are achieved thanks to three remotely controlled linear translation stages with micrometric precision mounted in a 3D configuration.

\begin{figure}[htbp]
\centering
\includegraphics[width=.7\textwidth]{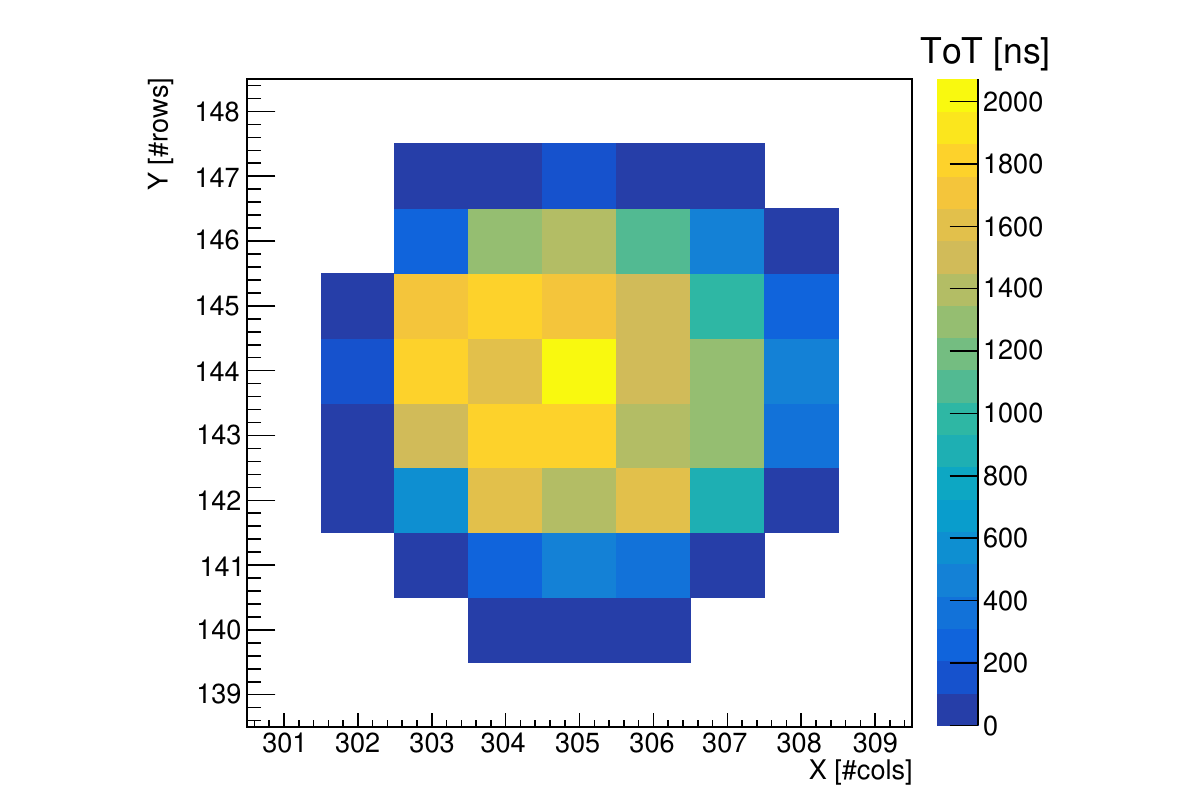}
\caption{ToT distribution among the pixel on a laser pulse event.\label{fig:laser_cluster}}
\end{figure}

After focusing, the laser spot has a gaussian shape with a standard deviation measured to about 1.4 pixels, as shown in Figure~\ref{fig:laser_cluster}. In this way, a wide Time-over-Threshold range has been covered.
The laser has been focused on a pixel close to the center of the bottom half matrix, illuminating the same region of pixels during the entire measurement.

In order to estimate the timing resolution contribution due to the laser setup and the pulse generator alone, the jitter of the laser pulse, triggered by one output of the pulse generator, has been measured at the oscilloscope (Tektronix MSO 70804C, $8\;\giga\hertz$, $25\;\giga\textrm{S}\per\second$) with respect to the second output of the pulse generator (that is usually sent to a Timepix4 digital pixel for the other measurements presented in this work). The laser pulse has been measured with an Ultrafast Photodiode (ALPHALAS InGaAs Photodetector UPD-35-UVIR-P).
The measured jitter amounts to $(10.9\pm0.1)\;\pico\second$ r.m.s., which is negligible compared to the other contributions to the detector timing resolution, as described in section \ref{sec:timing_res_contributions}.

\section{Measurements and analysis description}\label{sec:analysis}

The typical measurement consists of illuminating the Timepix4 assembly in a fixed position with a variable amount of light per pulse, at a fixed repetition rate ($200\;\hertz$).

The Timepix4 output data received by the data acquisition PC are decoded, assigning to each event a spatial coordinate (X and Y), a timestamp (ToA) and a Time-over-Threshold value (ToT).
\newline Afterwards, the decoded events are clustered using a common Density Based algorithm (Density-Based Spatial Clustering of Applications with Noise or DBSCAN), which consists of searching for hits which neighbour is a seed hit in space and in time, thereby forming clusters \cite{bib:DBSCAN}. 

Once each cluster has been created, the ToA of each pixel in the cluster is compared with the ToA of the reference signal ($\textrm{ToA}_{\textrm{ref}}$), which is the one sent to the digital pixel input.

A priori each pixel in the cluster is expected to have a different time walk contribution, and different ToT vs charge (Q) calibration.

For this reason, the analysis has been performed initially on single pixels. However, a number of corrections and calibrations still have to be performed in order to obtain the best performance, as shown in the next sections. After all corrections and calibrations have been applied, cluster-related quantities, like cluster time and cluster charge, have been introduced.

\section{Procedure for precise VCO calibration }\label{sec:VCO_calibration}

\begin{figure}[htbp]
\centering
\includegraphics[width=.9\textwidth]{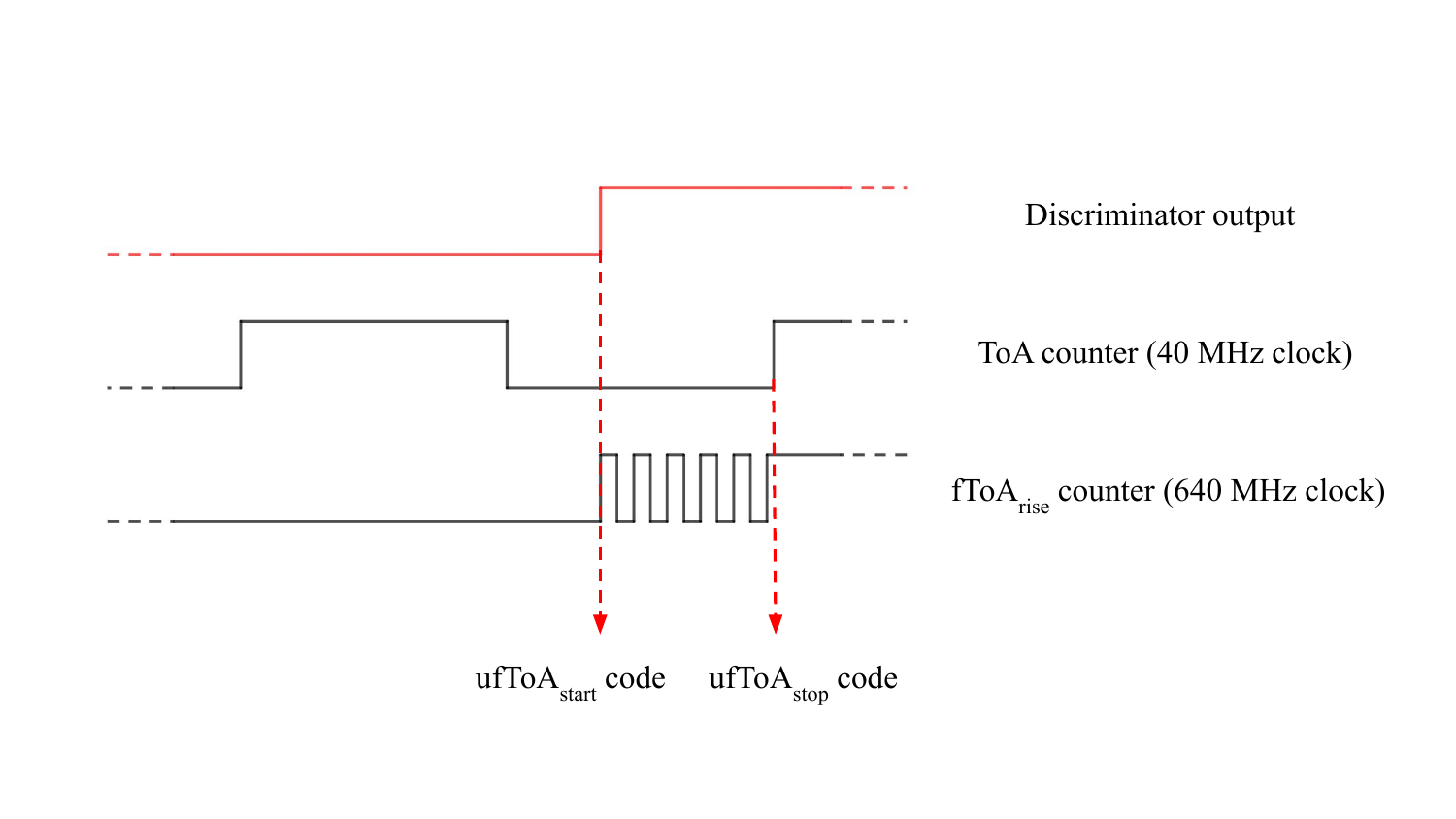}
\caption{Simplified clock schematic of the $\textrm{fToA}_{\textrm{rise}}$ and ufToA measurements. \label{fig:clock_fToA_rise}}
\end{figure}

During the measurements it has been observed that the different VCOs were oscillating at different frequencies.
This effect can be quantified by studying the length of a full fToA cycle: 
\begin{equation}
    \textrm{ufToA}_{\textrm{cycle}}=\textrm{fToA}_{\textrm{rise}}\cdot 8 + \textrm{ufToA}_{\textrm{stop}} - \textrm{ufToA}_{\textrm{start}}
    \label{eq:UfToA_cycle_definition}
\end{equation}
where $\textrm{fToA}_{\textrm{rise}}$, $\textrm{ufToA}_{\textrm{start}}$ and $\textrm{ufToA}_{\textrm{stop}}$ are the fine and ultra-fine corrections of the ToA described in paragraph \ref{subsec:Time_measurement}. Figure \ref{fig:clock_fToA_rise} represents a simplified schematic of the $\textrm{fToA}_{\textrm{rise}}$ and ufToA codes measurements. The $\textrm{fToA}_{\textrm{rise}}$ is the number of $640\;\mega\hertz$ clock cycles between the discriminator output rising edge and the following $40\;\mega\hertz$ clock cycle rising edge. The $\textrm{ufToA}_{\textrm{start}}$ and $\textrm{ufToA}_{\textrm{stop}}$ codes are obtained as shown in Figure \ref{fig:UfToA_operation}, respectively from the discriminator output rising edge and the following $40\;\mega\hertz$ clock cycle rising edge.

\begin{figure}[htbp]
\centering
\includegraphics[width=.7\textwidth]{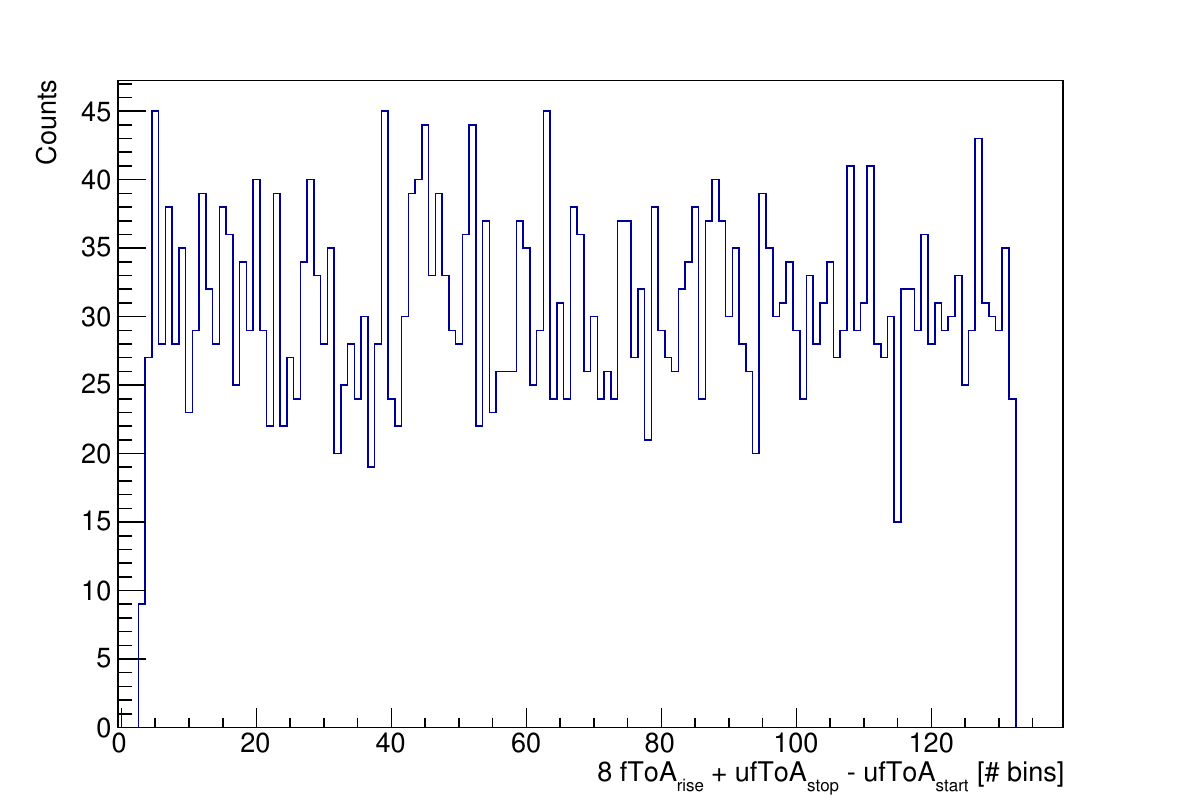}
\caption{$\textrm{ufToA}_{\textrm{cycle}}$ distribution obtained measuring laser pulses on a single pixel. \label{fig:UfToA_cycle_histo}}
\end{figure}

The quantity defined in Equation \eqref{eq:UfToA_cycle_definition} represents the number of ultra-fine bins that would fit between the signal rising edge and the next reference clock rising edge.
\newline The range of $\textrm{ufToA}_{\textrm{cycle}}$ is obtained for each TDC by plotting a histogram of the values measured when the pixel is triggered with externally applied digital test pulses which are not synchronised to the $40\;\mega\hertz$ clock, as shown in Figure \ref{fig:UfToA_cycle_histo}.

Knowing that this range would always cover a period of $25\;\nano\second$, depending only on the stability of the $40\;\mega\hertz$ clock, in nominal conditions, i.e. the VCO oscillating at $\nu_{\textrm{VCO}}=640\;\mega\hertz$, this range would be
\begin{equation}
    \textrm{range}(\textrm{ufToA}_{\textrm{cycle}})=8\cdot \textrm{nbins}(\textrm{fToA}) = 8\cdot \frac{\nu_{\textrm{VCO}}}{\nu_{\textrm{refclk}}} = 8\cdot\frac{640\;\mega\hertz}{40\;\mega\hertz} = 128
    \label{eq:UfToA_cycle_example}
\end{equation}

Observing the histogram for different pixels not sharing a common VCO, it is possible to see that different VCOs oscillate at different frequencies. 
\newline If the frequency is not correctly taken into account both the ufToA and the fToA bins width would be badly estimated. This turns into non-negligible errors when the rising edge is far enough from the next reference clock rising edge (when the VCO would stop), since there would be a large amount of fToA and ufToA bins contributing, and the bins width error would have to be considered for all of them. 
\newline As an example, if the range of $\textrm{ufToA}_{\textrm{cycle}}$ is 131 instead of 128, inverting Equation \eqref{eq:UfToA_cycle_example}, the resulting frequency of the VCO is $655\;\mega\hertz$, which turns into a ufToA bin width of $191\;\pico\second$ instead of $195\;\pico\second$. A signal measured on the $\textrm{ufToA}_{\textrm{cycle}}$ bin number 3 would be measured with a decoded ToA mistaken by just $\epsilon_3 = \textrm{ToA}_3 - \textrm{ToA}_3^* = 3\cdot 4\;\pico\second = 12\;\pico\second$, but the error on a signal measured on the bin 130 would be $\epsilon_{131} = \textrm{ToA}_{131} - \textrm{ToA}_{131}^* =520\;\pico\second$.

\medskip
Therefore, it is mandatory to measure and calibrate the frequency for the TDCs in the whole matrix.
To perform this calibration, internal digital test pulses have been sent to one pixel per each TDC to measure the VCO frequency.

\begin{figure}[htbp]
\centering
\includegraphics[width=.7\textwidth]{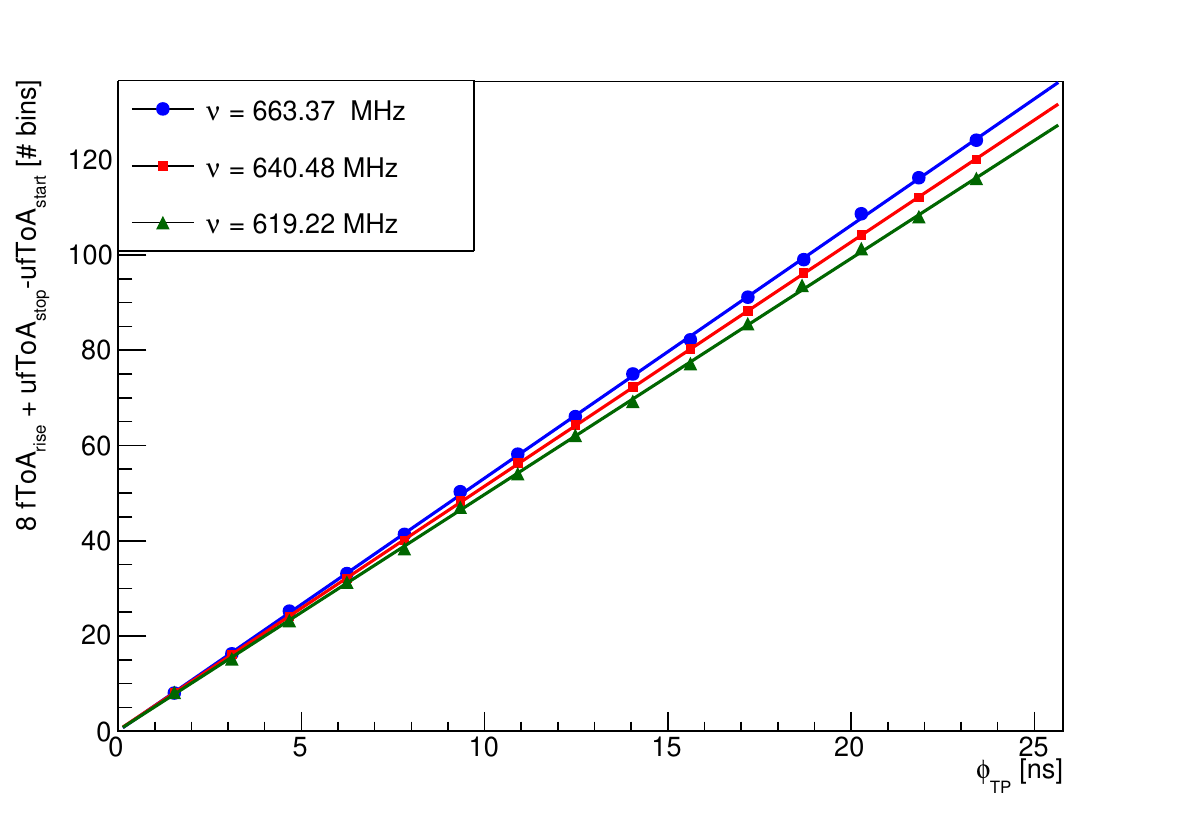}
\caption{$\textrm{ufToA}_{\textrm{cycle}}$ dependence on VCO for different pixels. \label{fig:fit_VCO_calibration}}
\end{figure}

\medskip
An internal test pulse synchronized with the $40\;\mega\hertz$ reference clock is available, and it is possible to set its phase $\phi$ with respect to the reference clock. Fixing the phase with respect to the clock, the populated $\textrm{ufToA}_{\textrm{cycle}}$ bin is always the same. Varying the phase, different bins are populated, defining pairs $(\phi,\textrm{ufToA}_{\textrm{cycle}})$.
\newline The 16 pairs $(\phi,\textrm{ufToA}_{\textrm{cycle}})$ have then been fitted with a linear function $\textrm{ufToA}_{\textrm{cycle}} = \textrm{p}_0 + \textrm{p}_1\cdot\phi$, as shown in Figure 
\ref{fig:fit_VCO_calibration}, whose slope $\textrm{p}_1\;\left[\hertz\right]$ is associated to the VCO frequency through the relation
\begin{equation}
    \nu_{\textrm{VCO}} = \frac{\textrm{p}_1}{8}
\end{equation}
where the coefficient 8 is the ratio between the width of the fToA bins and the one of the ufToA bins.

\begin{figure}[htbp]
\centering
\subfloat[]
{
    \centering
    \includegraphics[width=.95\textwidth]{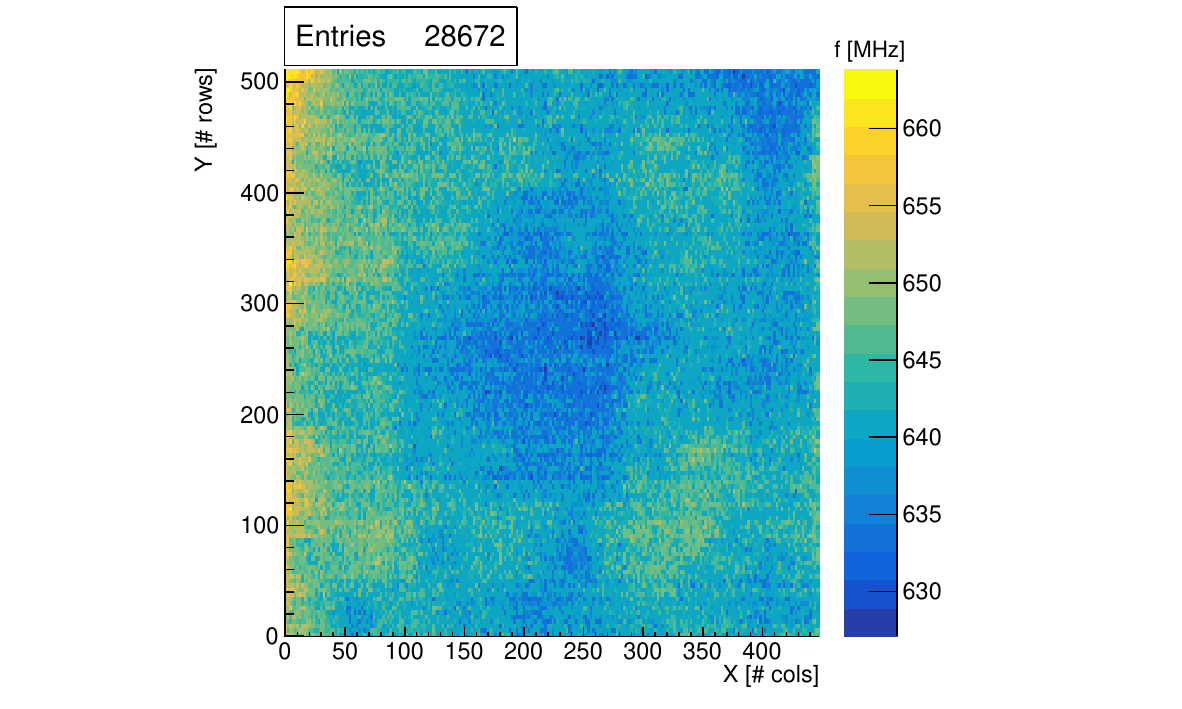}
    \label{fig:VCO_calibration_2d}
}
\hfill
\subfloat[]
{
    \centering
    \includegraphics[width=.72\textwidth]{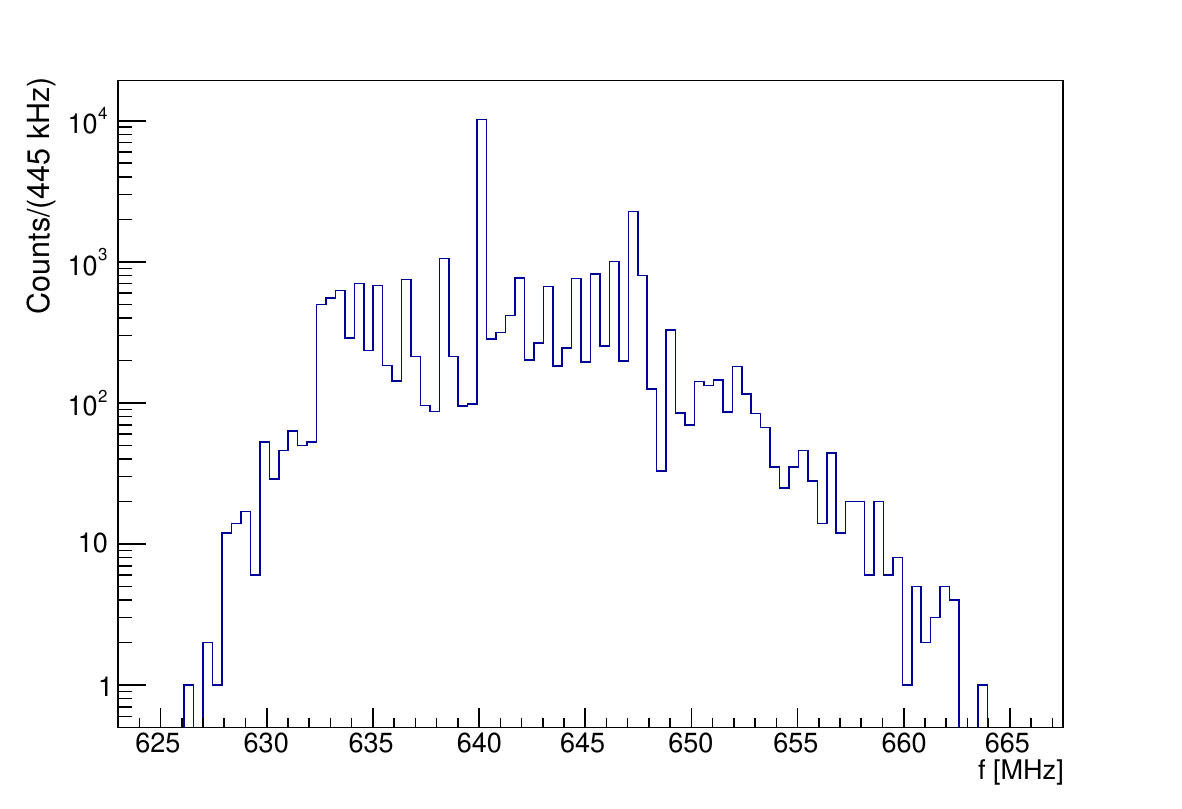}
    \label{fig:VCO_calibration_1d}
}
\caption{\textbf{(a)} VCO frequency bi-dimensional distribution over the whole matrix. \textbf{(b)} VCO frequency values distribution.\label{fig:VCO_calibration}}
\end{figure}

\medskip
By measuring this frequency for each TDC a map of the VCO frequencies has been produced, as shown in Figure \ref{fig:VCO_calibration_2d}. 

In Figure \ref{fig:VCO_calibration_1d}, which is an histogram of the values in Figure \ref{fig:VCO_calibration_2d}, it can be noticed that the frequency distribution is peaked at the nominal frequency $640\;\mega\hertz$, but it covers a range of $\sim40\;\mega\hertz$; the circuit allows to adjust the VCO frequency on a range of $\sim20\;\mega\hertz$, so it would not be possible to perform measurements adjusting all the VCOs in order to make them run at the same frequency. However, this will not be an issue, since correcting their contribution on the offline analysis would preserve the timing resolution, as described in subsection {\ref{subsec:VCO_frequency}}.
\newline The Timepix4 on-pixel TDC's VCO frequency is locked by a PLL running with the same VCO placed at the center of the chip. The distributed control voltage suffers from local GND reference mismatch due to internal power supply drops. Figure \ref{fig:VCO_calibration} is therefore is an indirect measurement of the internal digital power supply which corresponds approximately to $\sim 1\;\milli\volt\per\mega\hertz$.

\section{Time-over-Threshold calibration}\label{subsec:Q_vs_ToT_calib}

\begin{figure}[htbp]
\centering
\includegraphics[width=.7\textwidth]{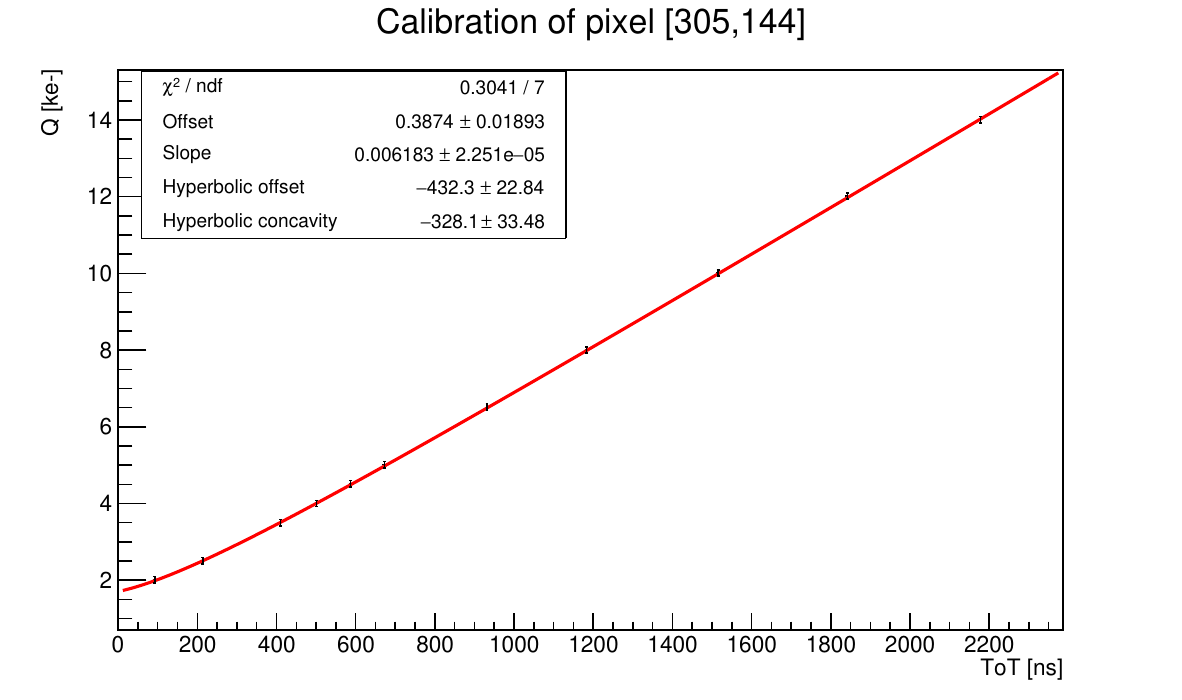}
\caption{Example of the charge non-linear calibration against ToT of a single pixel. \label{fig:Single_pixel_calibration}}
\end{figure}

A per pixel calibration of the ToT measured against the deposited charge has been performed on the whole matrix, using the Timepix4 internal analog test pulse, that is a signal with known and selectable amplitude injected in a capacitor placed at the input of the front-end chain. This allows to inject on each pixel several pulses carrying different charges. The capacitances can vary by about $1\;\%$ in different pixels, but for the calibration they have been assumed as equal. This may slightly degrade the calibration performance.
The ToT depends non-linearly on the deposited charge, so for each pixel the pairs $\left(\textrm{ToT}, \textrm{Q}\right)$ have been fitted using the function:
\begin{equation}
    \textrm{Q}(\textrm{ToT})=\textrm{p}_0 + \textrm{p}_1\cdot \textrm{ToT} - \frac{\textrm{p}_2}{\textrm{ToT}-\textrm{p}_3}
\end{equation}
where Q is in $\kilo\textrm{e}^-$ and ToT is in ns. A plot showing the measured ToT calibration for a pixel is shown in Figure \ref{fig:Single_pixel_calibration}, with superimposed fitting function. The maximum injected charge corresponds to about $14\;\kilo\textrm{e}^-$ due to the test pulse saturation for the particular DAC parameters used in this measurement, presented in Table \ref{tab:DAC_settings}.

\begin{figure}[htbp]
\centering
\includegraphics[width=.9\textwidth]{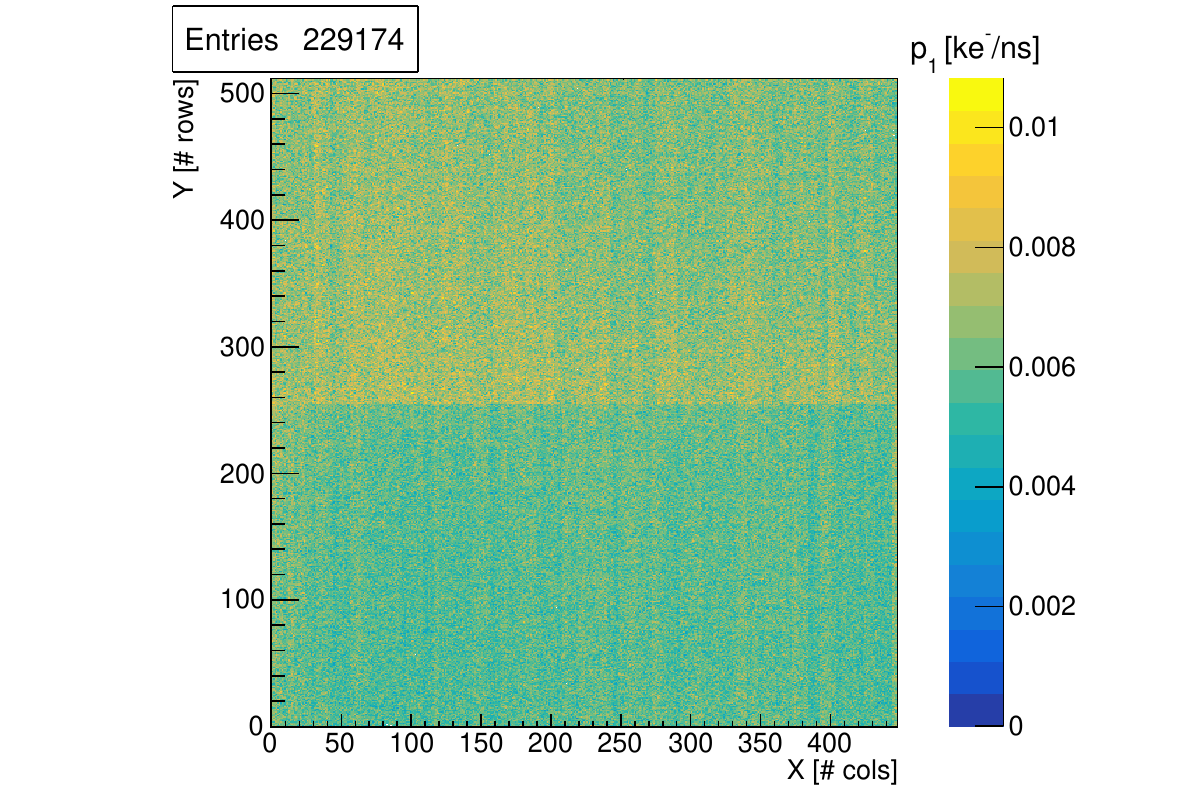}
\caption{Example of the fit parameter $\textrm{p}_1$ distribution over the matrix. \label{fig:slope_calibration_distribution}}
\end{figure}

By repeating this procedure on all Timepix4 pixels ($\sim230$ thousands) a bi-dimensional distribution of the fit parameters has been obtained, allowing a per-pixel calibration covering the whole matrix. As an example, Figure \ref{fig:slope_calibration_distribution} presents the distribution across the matrix of the $\textrm{p}_1$ parameter. A structure due to the gain difference between top and bottom matrix is clearly visible, as well as vertical structures due to different behaviour of different double columns.

\begin{figure}[htbp]
\centering
\subfloat[]
{
    \centering
    \includegraphics[width=.48\textwidth]{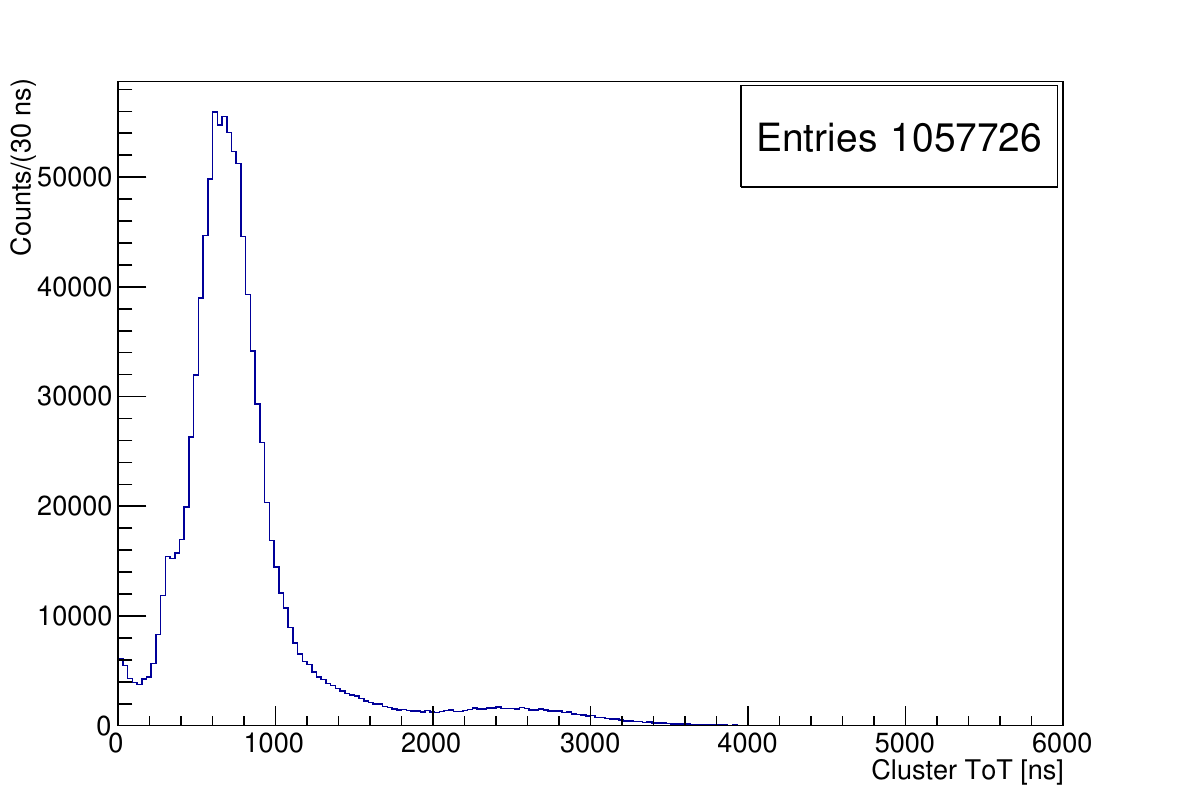}
    \label{fig:Am_Cs_not_calibrated}
}
\hfill
\subfloat[]
{
    \centering
    \includegraphics[width=.48\textwidth]{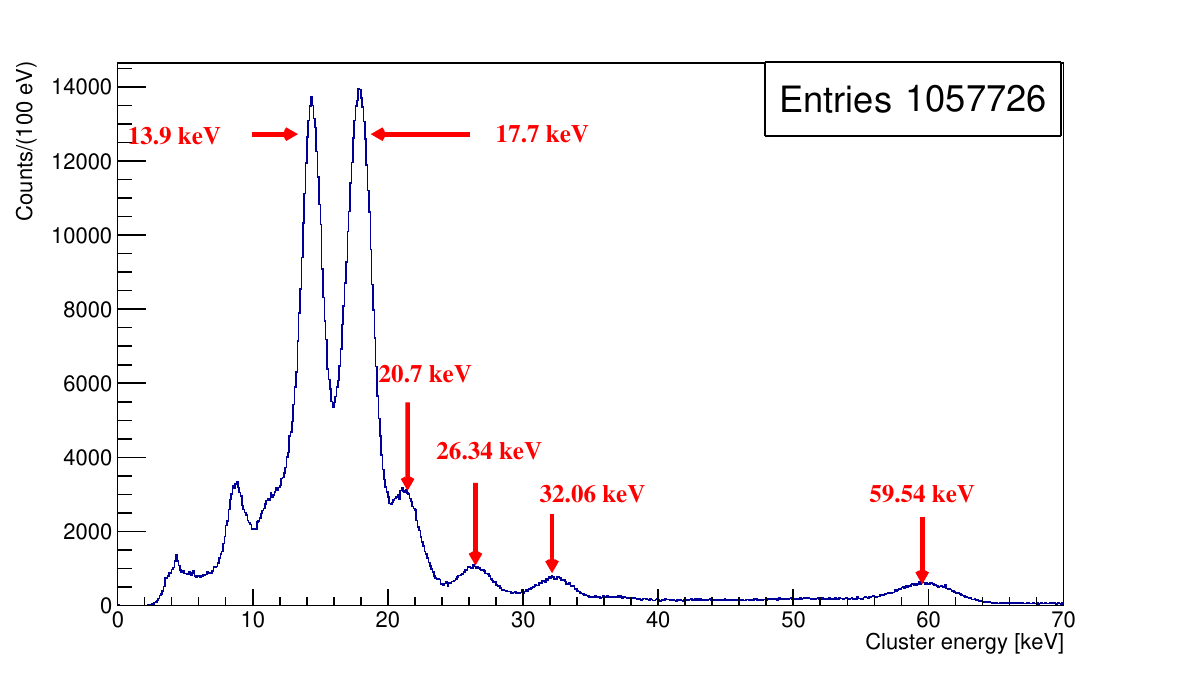}
    \label{fig:Am_Cs_calibrated}
}
\hfill
\subfloat[]
{
    \centering
    \includegraphics[width=.48\textwidth]{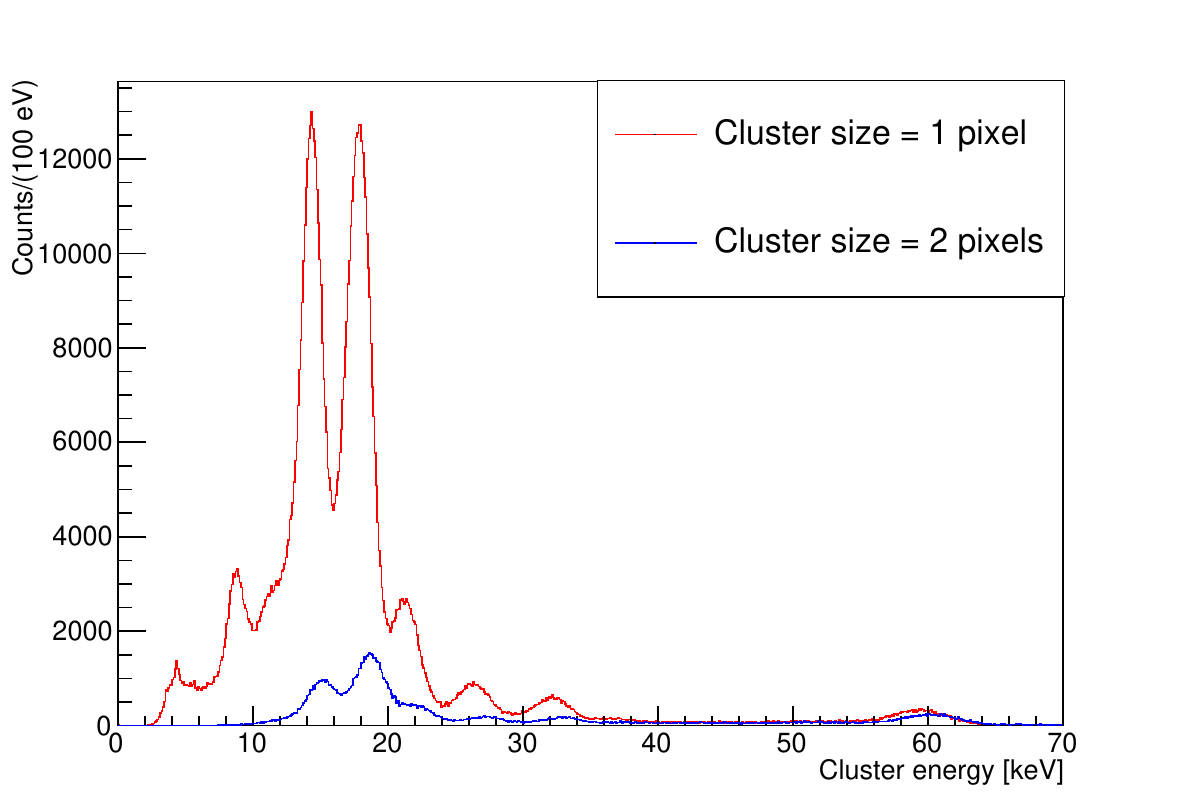}
    \label{fig:Am_Cs_calibrated_size_comparison}
}
\hfill
\subfloat[]
{
    \centering
    \includegraphics[width=.48\textwidth]{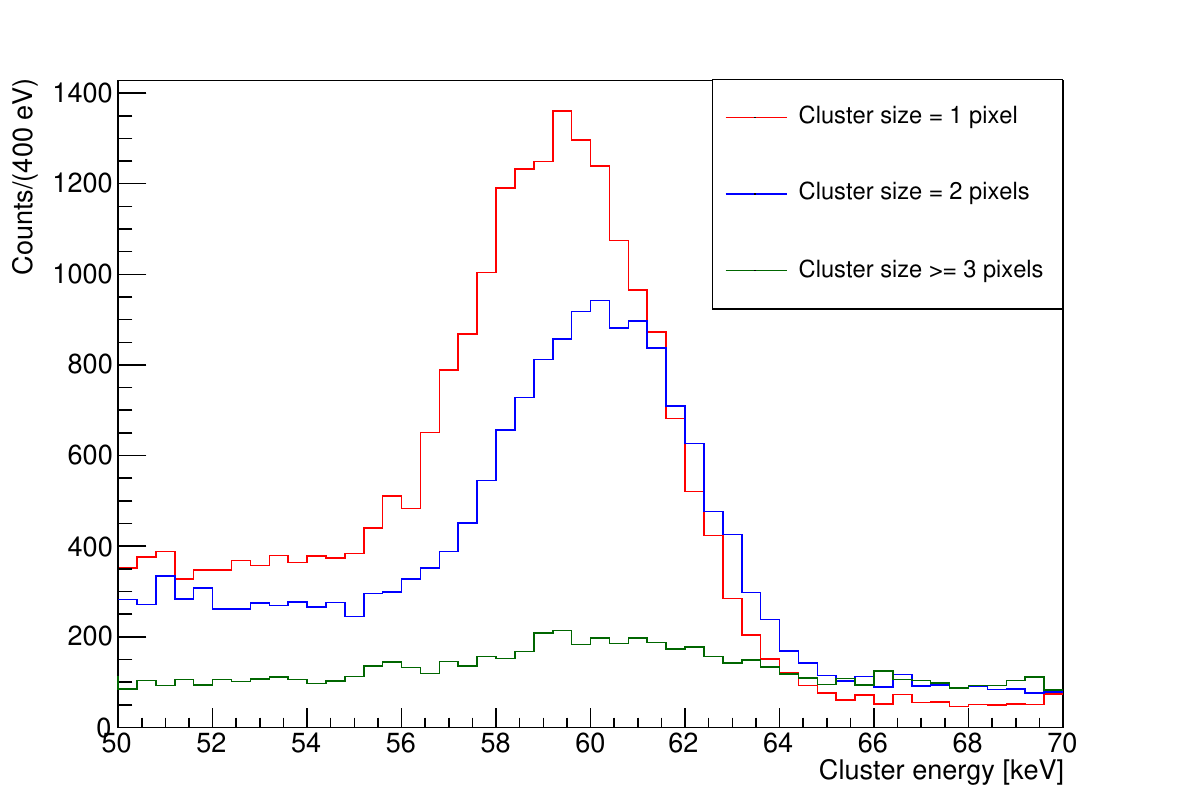}
    \label{fig:Am_Cs_calibrated_size_comparison_60keV}
}
\hfill
\subfloat[]
{
    \centering
    \includegraphics[width=.67\textwidth]{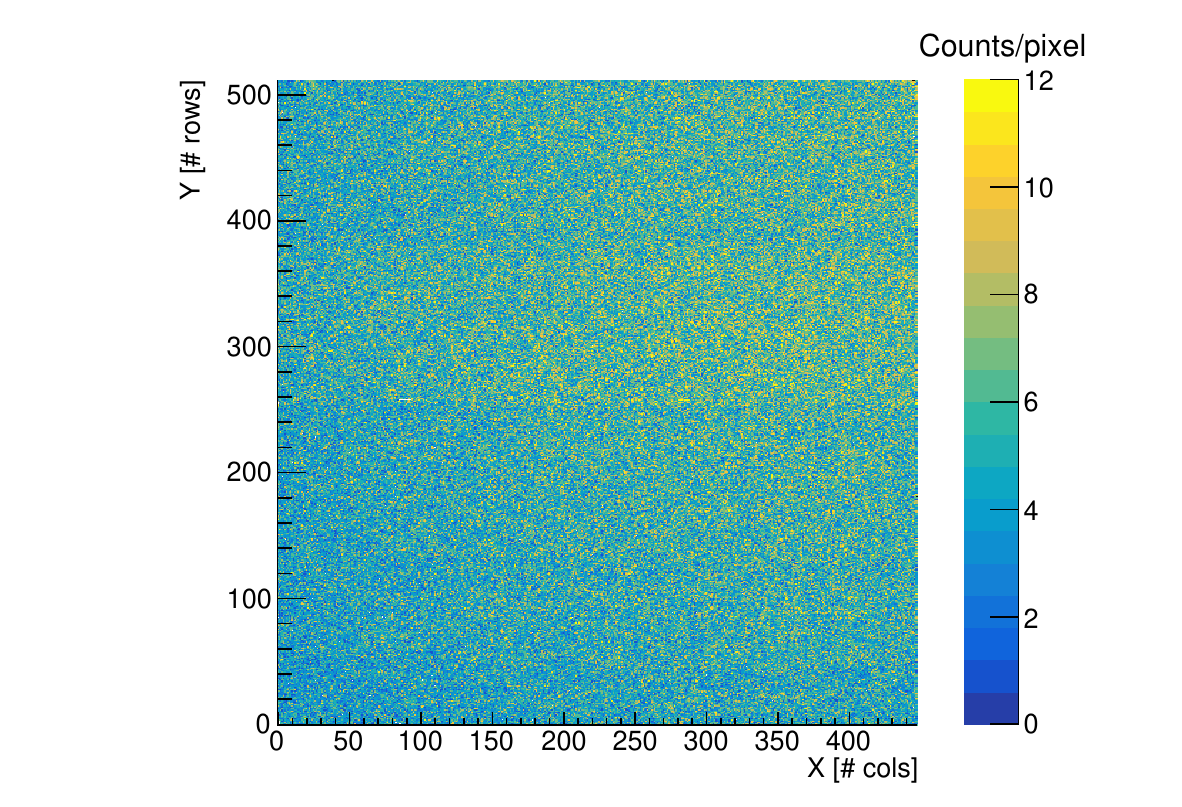}
    \label{fig:hitmap_source}
}
\caption{\textbf{(a)} $^{241}\textrm{Am}$ and $^{137}\textrm{Cs}$ ToT spectrum measured on the whole matrix before the per-pixel calibration, considering all clusters. \textbf{(b)} $^{241}\textrm{Am}$ and $^{137}\textrm{Cs}$ energy spectrum measured on the whole matrix after the per-pixel calibration, considering all clusters. \textbf{(c)} Comparison of the calibrated spectrum due to 1-pixel clusters and 2-pixels clusters. \textbf{(d)} Comparison of the peak at $59.54\;\kilo\electronvolt$ resulting from clusters with size 1, 2 or greater. \textbf{(e)} Hitmap of the events measured irradiating the matrix with $^{137}\textrm{Cs}$ and $^{241}\textrm{Am}$. \label{fig:Am_Cs_spectrum}}
\end{figure}

The resulting calibration has been validated measuring the X-rays spectrum produced by $^{241}\textrm{Am}$ and $^{137}\textrm{Cs}$ sources used to simultaneously irradiate the whole detector matrix from a few cm distance. All the pixels, except for the noisy ones, have been enabled and irradiated, as shown in Figure \ref{fig:hitmap_source}. The injected charge expressed in $\kilo\textrm{e}^-$ is converted in $\kilo\electronvolt$ when using the sources by multiplying by a factor $3.6\;\electronvolt\per\textrm{e}^-$, which is the average energy to produce $e-h$ pairs in Si. The improvement given by the per pixel calibration to the resolving power is clearly visible in the difference between the uncalibrated cluster ToT spectrum (Figure \ref{fig:Am_Cs_not_calibrated}) and the energy calibrated one (Figure \ref{fig:Am_Cs_calibrated}). The $^{241}\textrm{Am}$ peaks at $13.9\;\kilo\electronvolt$, $17.8\;\kilo\electronvolt$, $20.8\;\kilo\electronvolt$, $26.35\;\kilo\electronvolt$ and $59.5\;\kilo\electronvolt$ are well resolved and their energy is correctly estimated, as well as the $^{137}\textrm{Cs}$ peak at $32.0\;\kilo\electronvolt$.
Figures \ref{fig:Am_Cs_calibrated_size_comparison} and \ref{fig:Am_Cs_calibrated_size_comparison_60keV} show a spectrum comparison with respect to the cluster size considered, respectively for the whole spectrum up to $70\;\kilo\electronvolt$ and only for the peak at $59.54\;\kilo\electronvolt$. An offset between the peaks related to single pixel clusters and the ones related to two-pixels clusters is present, due to a small calibration error introduced by the uncertainty on the capacitors used to convert the test pulse DAC voltage into a charge on each pixel.

In Table \ref{tab:peaks_energy} the energy values corresponding to the radioactive source measurements are summarised: the expected energies of the measured peaks, together with the measured ones and the full-width-half-maximum values of the Gaussian fit used to extract these measurements.

\begin{table}[!ht]
\centering
\begin{tabular}{|c|c|c|c|}
\hline
\textbf{Nuclide} & $\mathbf{E_{expected} [\kilo\electronvolt]}$ & $\mathbf{E_{measured} [\kilo\electronvolt]}$ &  $\mathbf{FWHM_{measured} [\kilo\electronvolt]}$ \\ \hline
 $\textrm{Np}\;\textrm{L}\alpha$ & 13.9 & 14.3 &  1.6 \\ \hline
 $\textrm{Np}\;\textrm{L}\beta_1\;+\;\textrm{Np}\;\textrm{L}\beta_2 $ & 17.7 & 17.8 &  2.0 \\ \hline
 $\textrm{Np}\;\textrm{L}\gamma_1$ & 20.7 & 21.3 & 2.4 \\ \hline
 $^{241}\textrm{Am}$ & 26.3 & 26.5 &  2.6 \\ \hline
 $\textrm{Ba}\;\textrm{K}\alpha_1+\textrm{K}\alpha_2$ & 32.0 & 32.0 &  2.7 \\ \hline
 $^{241}\textrm{Am}$ & 59.5 & 59.4 &  4.7 \\ \hline
\end{tabular}
\caption{Measured energy of $^{137}\textrm{Cs}$ and $^{241}\textrm{Am}$ main emissions compared with expected energies, considering all clusters.}
\label{tab:peaks_energy}
\end{table}

\section{Timing resolution contributions}\label{sec:timing_res_contributions}

The achievable timing resolution obtained from laser illumination is affected by several contributions, described in this section.
Some of these contributions, i.e. the TDC resolution, are intrinsic and cannot be removed, while others, i.e. the time walk effect or the VCO frequency variations, can be corrected for. 

The impact of effects like time walk and jitter can be reduced by maximising the charge collected in a single pixel. The measurements presented in this section are performed under this assumption. The measurements are based on the central (most illuminated) pixel in the cluster. The laser attenuation was set in order to inject a charge included in $\left[50\;\kilo\textrm{e}^-;60\;\kilo\textrm{e}^-\right]$ in the central pixel, in order to minimise time walk and jitter.
\newline Further measurements performed with a variable attenuation will be described in Sections~\ref{sec:Single_pixel_timing_res} and \ref{sec:Cluster_timing_res}, together with a description of the time walk correction.

\subsection{Measured variation of the Voltage Controlled Oscillator frequency }\label{subsec:VCO_frequency} 

\medskip
Precise timing measurements require to take into account and correct for the different VCO frequencies, either in the Timepix4 configuration or in the offline analysis, as demonstrated in Section \ref{sec:VCO_calibration}.

\begin{figure}[htbp]
\centering
\includegraphics[width=.7\textwidth]{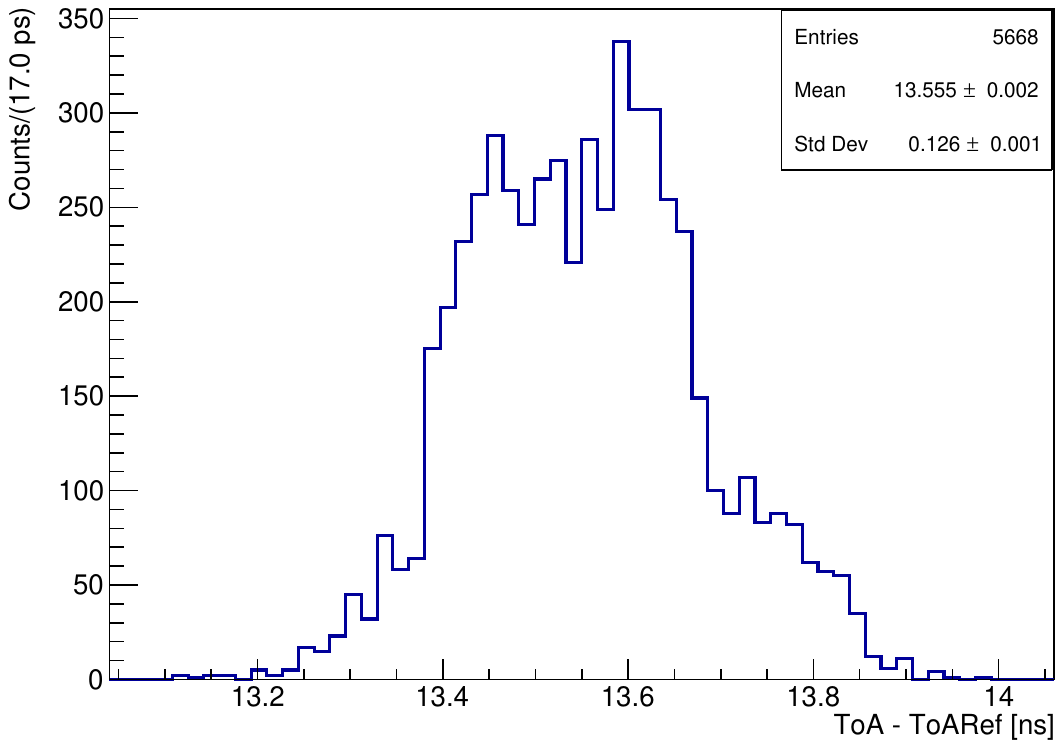}
\caption{Distribution of $\textrm{ToA}- \textrm{ToA}_{\textrm{ref}}$ after the VCO frequency correction. \label{fig:resolution_VCO_corrected}}
\end{figure}

Without taking into account VCO frequency variation, the measured timing resolution is about $200\;\pico\second$ r.m.s..
After correcting the timing measurements using the proper VCO oscillation frequency, the standard deviation of the ToA difference between the central pixel and the reference pixel is greatly improved:
\begin{equation}
    \sigma_{\textrm{VCO}}=126\pm1\;\pico\second\;\textrm{r.m.s.}
\end{equation} 
as shown in Figure \ref{fig:resolution_VCO_corrected}.

\medskip
The measurement above is based on a VCO frequency correction applied at the analysis level after data acquisition, during which the VCOs were running at different frequencies.

A second measurement has been performed by equalising the VCO frequencies at the level of the Timepix4 configuration, before the measurement, setting the VCO frequencies of the two pixels to the nominal $640\;\mega\hertz$ value. The results is $\sigma_{\textrm{adj}-\textrm{VCO}}=132\pm1\;\pico\second$ r.m.s., slightly worse than the previous measurement.

\medskip
The calibration of the VCOs frequencies, done so far only on the two pixels under test, has been extended to the whole matrix as described in Section~\ref{sec:VCO_calibration} to allow precise timing measurements on the whole Timepix4 active area. 
\newline All measurements presented in the following sections are corrected for the VCO frequency variation at the analysis level, after data acquisition.

\subsection{Reference signal resolution}\label{subsec:TDC_resolution}

\begin{figure}[htbp]
\centering
\includegraphics[width=.7\textwidth]{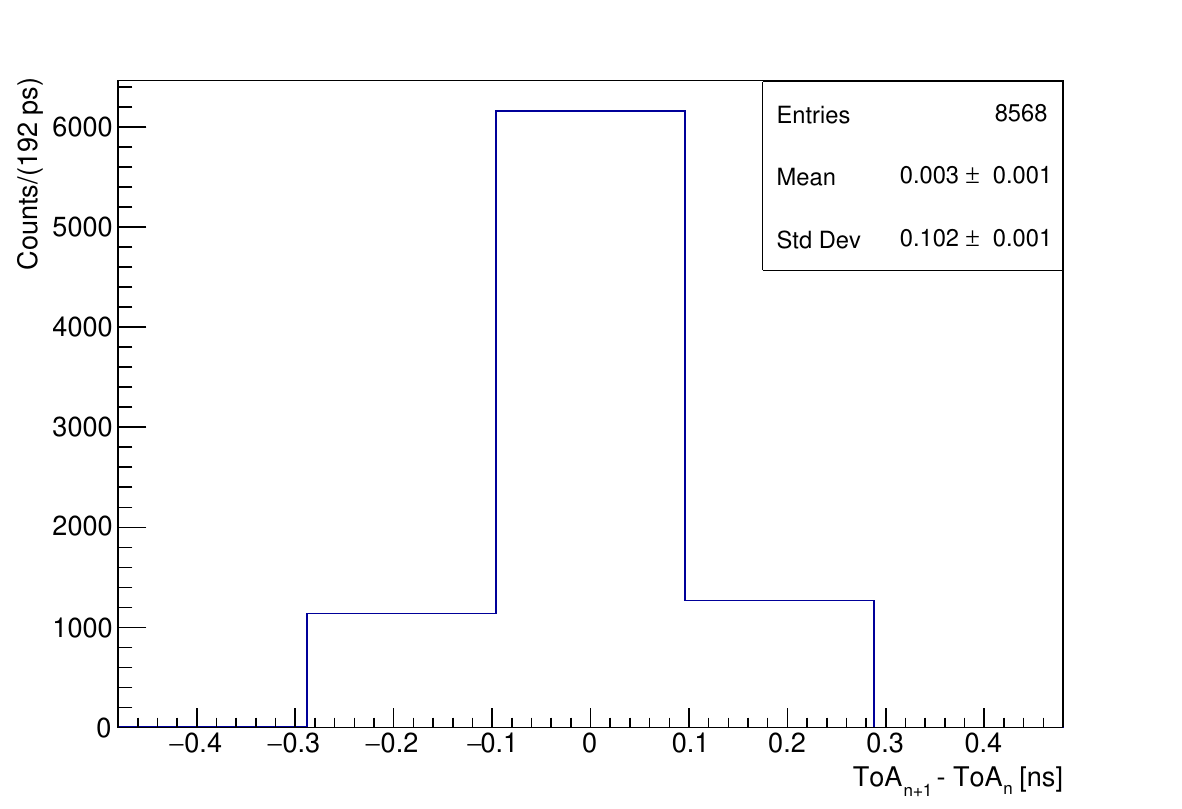}
\caption{Distribution of the period $\textrm{ToA}_{\textrm{n}+1}-\textrm{ToA}_{\textrm{n}}$ of external periodic test pulses sent to a single pixel through digital pixels. \label{fig:single_TDC_resolution}}
\end{figure}

The reference signal resolution has been estimated through externally generated test pulses.
A periodic pulse has been generated and sent to the pixel used as reference through the digital pixels, and the difference between the ToA of each pulse with the previous one has been computed. The external pulses are not synchronous to the Timepix4 reference clock at $40\;\mega\hertz$, and they are generated with a period ($\textrm{T} = (1165.000\pm0.007)\;\nano\second$) not multiple of $25\;\nano\second$, so they cover the entire range of fToA and ufToA bins. Thus, the resulting resolution already takes into account different bin widths.
The standard deviation of the $(\textrm{ToA}_{\textrm{n}+1} - \textrm{ToA}_{\textrm{n}})$ distribution amounts to $\sigma(\textrm{ToA}_{\textrm{n}+1} - \textrm{ToA}_{\textrm{n}})=102\pm1\;\pico\second$ r.m.s., as shown in Figure \ref{fig:single_TDC_resolution}. Being this result related to the difference between two signals registered by the same TDC, the resolution of the reference signal itself can be estimated, to a first approximation, as: 
\begin{equation}
    \sigma_{\textrm{REF}}=\sigma(\textrm{ToA}_{\textrm{n}+1}-\textrm{ToA}_{\textrm{n}})/\sqrt{2}=72\pm1\;\pico\second\;\textrm{r.m.s.}
\end{equation}

The above result is slightly higher than the resolution expected from a $195\;\pico\second$ time bin ($195\;\pico\second / \sqrt{12} = 56\;\pico\second$). This overestimation may be due to the contribution given by the generation of the external test pulses and their distribution to Timepix4 and within it. 

\subsection{Residual contributions to the timing resolution}\label{subsec:statistical_contribution}

The timing resolution obtained so far ($\sigma_{\textrm{VCO}}$) represents the error associated to the difference $\textrm{ToA}-\textrm{ToA}_{\textrm{ref}}$.
The timing resolution of a single Timepix4 pixel for a charge included in $\left[50\;\kilo\textrm{e}^-;\;60\;\kilo\textrm{e}^-\right]$ can therefore be calculated subtracting in quadrature the contribution due to the reference signal ($\sigma_{\textrm{REF}}$): 
\begin{equation}
    \sigma(\textrm{ToA})_{\textrm{final}}=\sqrt{\sigma_{\textrm{VCO}}^2-\sigma_{\textrm{REF}}^2} = 103\pm2\;\pico\second\;\textrm{r.m.s.}
\end{equation}

The value above includes, to a first approximation, the contributions due to the signal generation in the silicon sensor, the front-end jitter and the TDC bin width.

\section{Time walk correction and single pixel timing resolution charge dependence}\label{sec:Single_pixel_timing_res}

\begin{figure}[htbp]
\centering
\includegraphics[width=.7\textwidth]{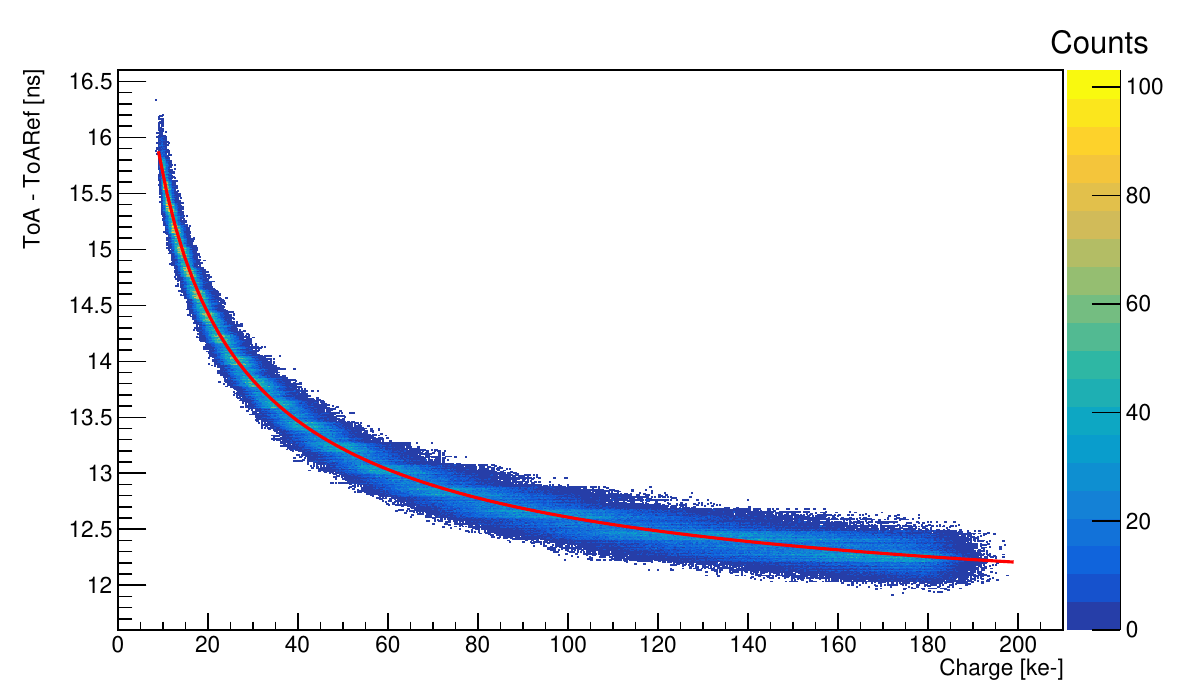}
\caption{Single pixel time walk distribution superimposed with the fitting function.\label{fig:time_walk_estimation}}
\end{figure}

To obtain precise timing measurements, it is necessary to correct for the time walk effect on single pixels. To estimate this effect, a set of measurements has been performed by varying the laser attenuation. In particular, the intensity was varied with fine steps. This allowed to cover charge values in the range of  $\left[\sim10\;\kilo\textrm{e}^-;\;200\;\kilo\textrm{e}^-\right]$ with a continuous-like distribution, and to obtain the time walk spectrum shown in Figure \ref{fig:time_walk_estimation}. As shown the distribution has been fitted using the function:
\begin{equation}
    \textrm{TW}(\textrm{ToT}) = \textrm{ToADiff}(\textrm{ToT})=\textrm{ToA}-\textrm{ToA}_{\textrm{ref}}=\frac{\textrm{p}_0}{\textrm{Q}^{\textrm{p}_1}+\textrm{p}_2} + \textrm{p}_3
    \label{eq:time_walk_fit}
\end{equation}

\begin{figure}[htbp]
\centering
\includegraphics[width=.7\textwidth]{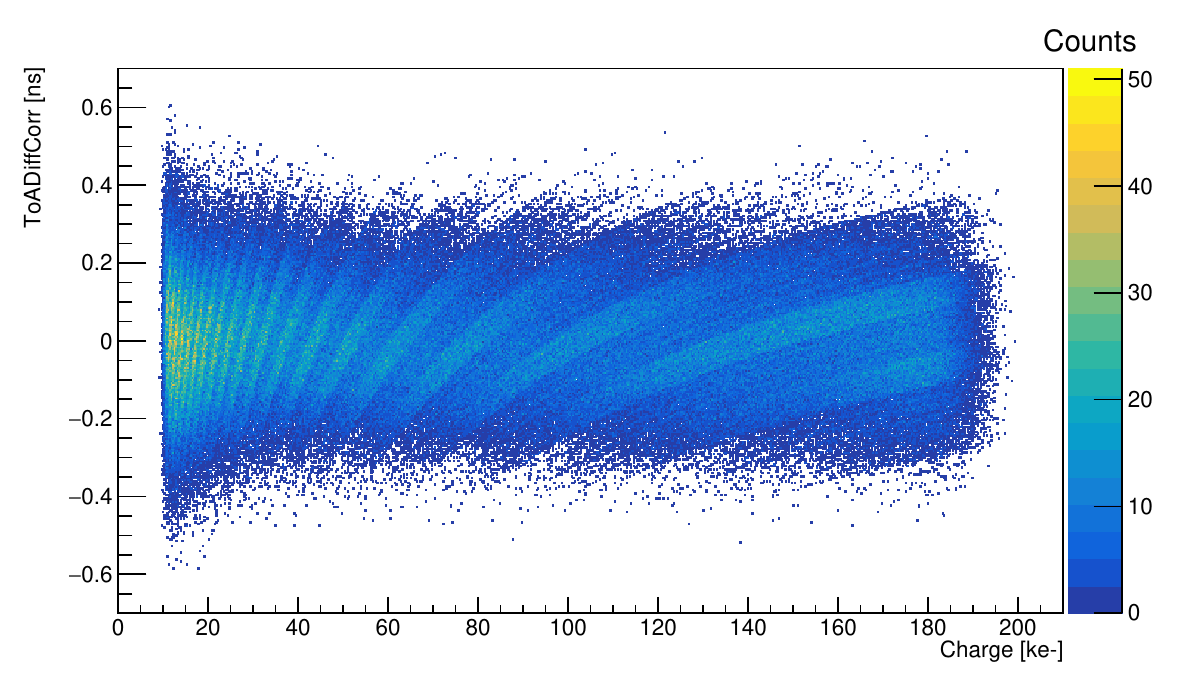}
\caption{Single pixel $\left(\textrm{Q},\textrm{ToA}-\textrm{ToA}_{\textrm{ref}}\right)$ distribution after time walk correction.\label{fig:time_walk_corrected}}
\end{figure}

Afterwards, the time walk effect has been corrected for each measured ToA value, by subtracting the charge-dependent correction given by Equation \eqref{eq:time_walk_fit}. The resulting time walk corrected plot is shown in Figure \ref{fig:time_walk_corrected}.

\begin{figure}[htbp]
\centering
\includegraphics[width=.7\textwidth]{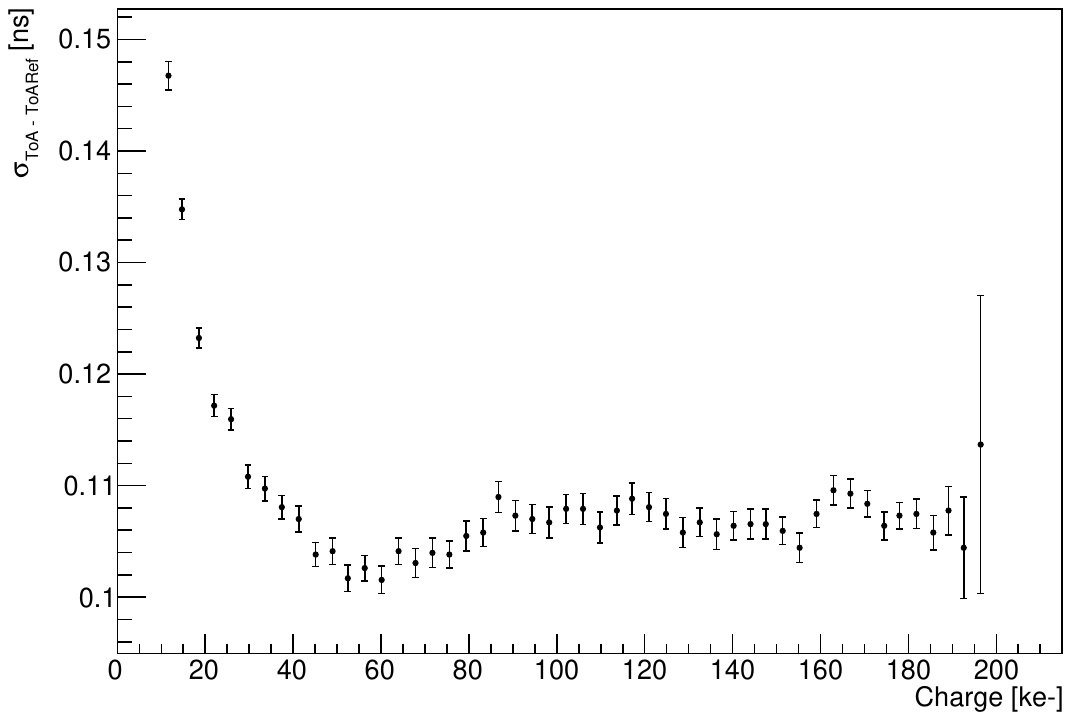}
\caption{Time resolution dependency with respect to the calibrated charge for a single pixel. Timing resolution of the reference pixel has been subtracted. \label{fig:time_resolution_single_pixel}}
\end{figure}

The distribution in Figure \ref{fig:time_walk_corrected} was divided in several "vertical slices", each one selecting a narrow charge range of $\sim4\;\kilo\textrm{e}^-$. 
The timing resolution has been estimated for each charge slice as the standard deviation of the distribution of the slice's projection on the ToADiffCorr axis.
The resulting resolutions are plotted in Figure \ref{fig:time_resolution_single_pixel}.

Starting from low input charges the values the timing resolution difference improves, reaching a value of $\sigma_{\textrm{ToA} - \textrm{ToARef}}=124.4\pm\ 0.8\;\pico\second\;\textrm{r.m.s.}$ for charge values $\textrm{Q} \in \left[52\;\kilo\textrm{e}^-;\;60\;\kilo\textrm{e}^-\right]$. At higher input charge values the timing resolution difference degrades again, reaching a saturation at $\sigma_{\textrm{ToA} - \textrm{ToARef}}=129\pm\ 2\;\pico\second\;\textrm{r.m.s.}$ for charge values higher than $80\;\kilo\textrm{e}^-$.
The single pixel timing resolution at $80\;\kilo\textrm{e}^-$ input charge is estimated by subtracting in quadrature the contribution due to the reference signal: this results to a single pixel timing resolution of $\sigma_{\textrm{ToA}}=107\pm 3\;\pico\second\;\textrm{r.m.s.}$.

The charge distribution (ToT) obtained when measuring using a laser with fixed intensity is gaussian-like, with a spread due both to the electronic noise introduced by the Timepix4 and the intrinsic spread given by the silicon detector. At fixed laser intensity, the timing distribution should also resemble a gaussian distribution. However, as the arrival time measurement is quantised into bins of $195\;\pico\second$ one might expect the hits to be spread over a few bins of $195\;\pico\second$, in the ideal case where frequencies of the reference VCO and the VCO of laser-excited pixel are identical. However in reality the oscillation frequency of the VCOs are slightly different and this leads to a finer quantisation, but still with peaks, as shown in Figure \ref{fig:resolution_VCO_corrected}. This timing quantisation generates the horizontal structures visible in Figure \ref{fig:time_walk_estimation}, which become tilted bands in Figure \ref{fig:time_walk_corrected} after subtracting the time walk contribution.

With higher injected charges, the spread of the measured charge is wide enough to cause an overlap of the distributions at different laser intensities. 

Due to these effects, the $\textrm{ToA} - \textrm{ToARef}$ distributions related to different laser intensities will overlap when selecting the events using the "vertical slice" method described above. This overlap and the distribution quantisation will result on a spread larger than the one that would be obtained performing a single measurement at a fixed charge corresponding to the considered slice charge. This explains the degradation of the resolution observed in Figure \ref{fig:time_resolution_single_pixel} at high charges ($\textrm{Q}\gtrsim70\;\kilo\textrm{e}^-$). 

The resulting resolution at input charges lower than $20\;\kilo\textrm{e}^-$ is slightly higher than the one expected taking into account the TDC contribution and the jitter of the preamplifier output when operating in electron collection mode \cite{bib:BALLABRIGA2023167489}. This difference can be attributed to an increased parasitic capacitance of the Si sensor and to the DAC settings reported in Table \ref{tab:DAC_settings} that affect the timing resolution at low input charges as discussed in Section \ref{sec:setup}. The increased sensor capacitance leads to a decrease in the slope of the preamplifier output rising edge, degrading its jitter. A further contribution to the degradation of the time resolution can be attributed to the electrons measured closed to the pixel edge instead of in its center which leads to position dependent shapes of the waveforms at the preamplifier output that cannot be compensated with the energy (ToT) correction of the time of arrival (ToA) \cite{bib:BALLABRIGA2023167489}.

\section{Cluster timing resolution}\label{sec:Cluster_timing_res}

When considering a cluster of more than one pixel, the information from the different pixels can be combined to extract more precise timing information, provided that the time walk effect at the single pixel level has been taken into account.

The ToA for each pixel in a cluster is corrected for the time walk effect using Equation \eqref{eq:time_walk_fit}. 
Afterwards, the ``cluster ToA'' ($\textrm{ToA}_{\textrm{cluster}}$) is calculated as the weighted-average of the ToA of individual pixels in the clusters. The used weights are the charge values for each pixel.
Instead, the ``cluster charge'' is simply computed as the sum of the individual pixels' charge. 

\begin{figure}[htbp]
\centering
\includegraphics[width=.7\textwidth]{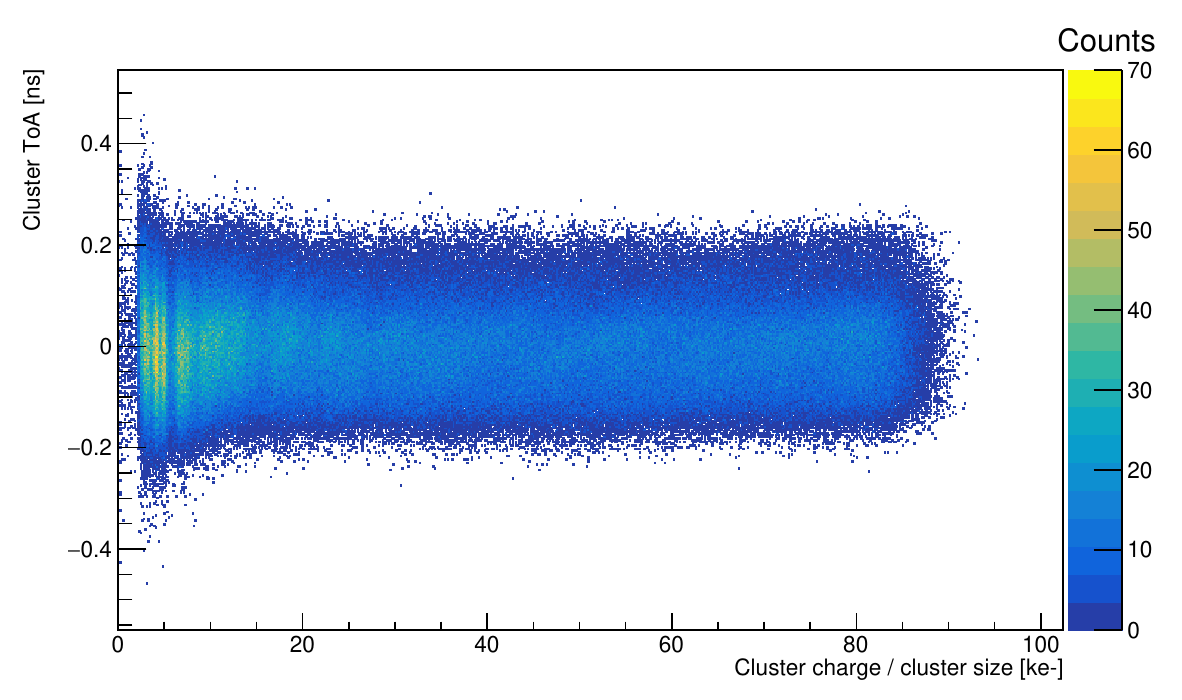}
\caption{Cluster ToA distribution as a function of the cluster charge. \label{fig:cluster_Charge_ToADiffAvg}}
\end{figure}

The cluster ToA as a function of the cluster charge  is shown in Figure \ref{fig:cluster_Charge_ToADiffAvg}.

\begin{figure}[htbp]
\centering
\includegraphics[width=.7\textwidth]{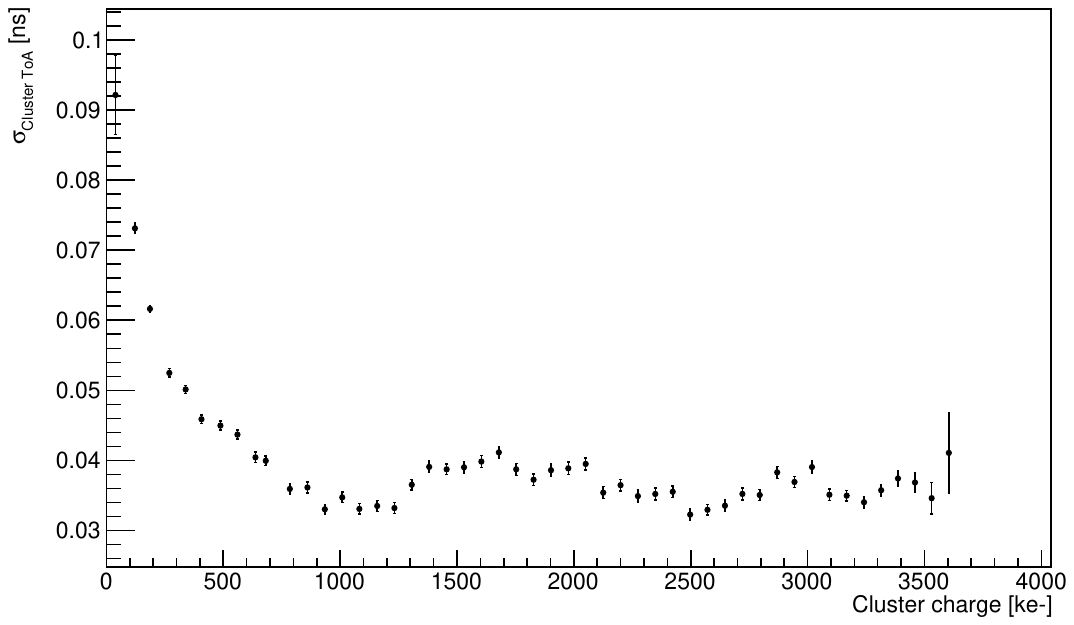}
\caption{Cluster timing resolution ($\sigma_{\textrm{ClusterToA}} = \sigma(\textrm{ToA}_{\textrm{cluster}}-\textrm{ToA}_{\textrm{ref}})$) as a function of the calibrated cluster charge. Timing resolution of the reference pixel has been subtracted. \label{fig:cluster_res_distribution}}
\end{figure}

The slice analysis described in Section~\ref{sec:Single_pixel_timing_res} was repeated over the complete cluster charge range. The distribution shown in Figure~\ref{fig:cluster_res_distribution}, where the ``cluster ToA'' resolution ($\sigma_{\textrm{ClusterToA}}$), evaluated as the standard deviation of the $\textrm{ToA}_{\textrm{cluster}}-\textrm{ToA}_{\textrm{ref}}$ distribution, is plotted as a function of the cluster charge.

As shown, the timing resolution distribution reaches a minimum for cluster charge $Q\in\left[\sim900\;\kilo\textrm{e}^-;\sim1300\;\kilo\textrm{e}^-\right]$, and the value at the minimum is:
\begin{equation}
    \sigma_{\textrm{ClusterToA}} = 79\pm 1\; \pico\second\;\textrm{r.m.s.}
\end{equation}

Once the reference signal contribution ($\sigma(\textrm{ToA}_{\textrm{ref}})$) is subtracted in quadrature, the minimum of cluster timing resolution distribution is:
\begin{equation}
    \sigma_{\textrm{True}\;\textrm{ClusterToA}} = 33 \pm 3\; \pico\second\;\textrm{r.m.s.}
\end{equation}

Measurement fluctuations of the order of a few $\pico\second$ in Figure~\ref{fig:cluster_res_distribution} are due to the intrinsic overlap among ToA distributions centered on different timing bins explained in Section \ref{sec:Single_pixel_timing_res}.

\subsection{Timing resolution dependence on the cluster size}\label{subsubsec:shells_timing_res}

The laser spot position onto the silicon sensor is fixed in all presented measurements, as shown in Figure \ref{fig:laser_cluster}. Around the central pixel, the most illuminated one and therefore the one with the largest charge value, there are a number of pixels which have an induced signal either due to the optical spread of the laser beam (and possible internal reflections) or charge sharing effects.

\begin{figure}[htbp]
\centering
\includegraphics[width=.9\textwidth]{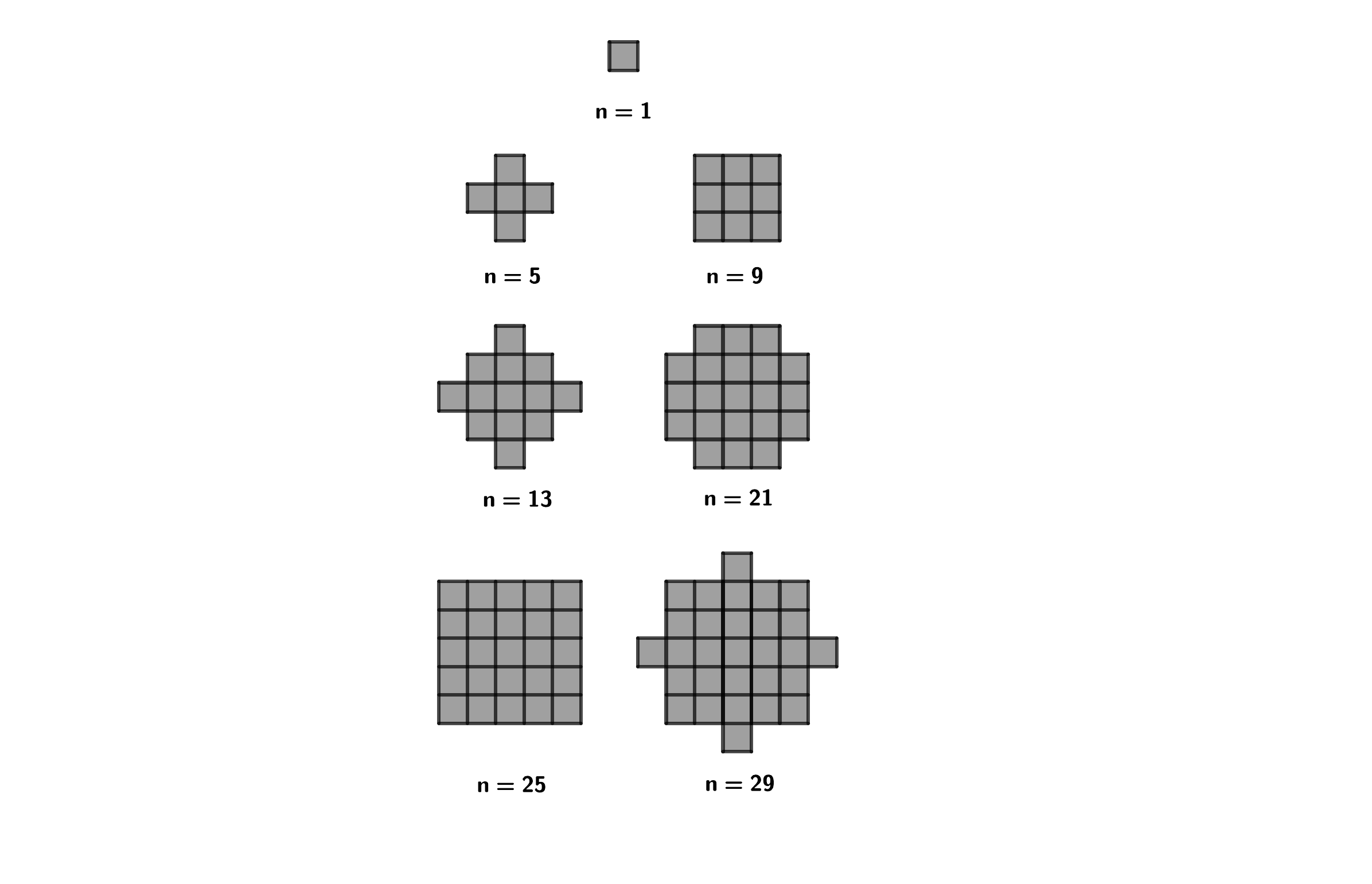}
\caption{Sketch showing clusters of different size (n) artificially created via software.\label{fig:shells}}
\end{figure}

It is important to estimate the minimum cluster size in order to reach the best resolution results obtained in Section \ref{sec:Cluster_timing_res}. To do that, the analysis described above has been repeated creating ``artificial'' clusters with different sizes, removing selected pixels via software from the  original physical cluster as shown in Figure \ref{fig:shells}.

\begin{figure}[htbp]
\centering
\includegraphics[width=.7\textwidth]{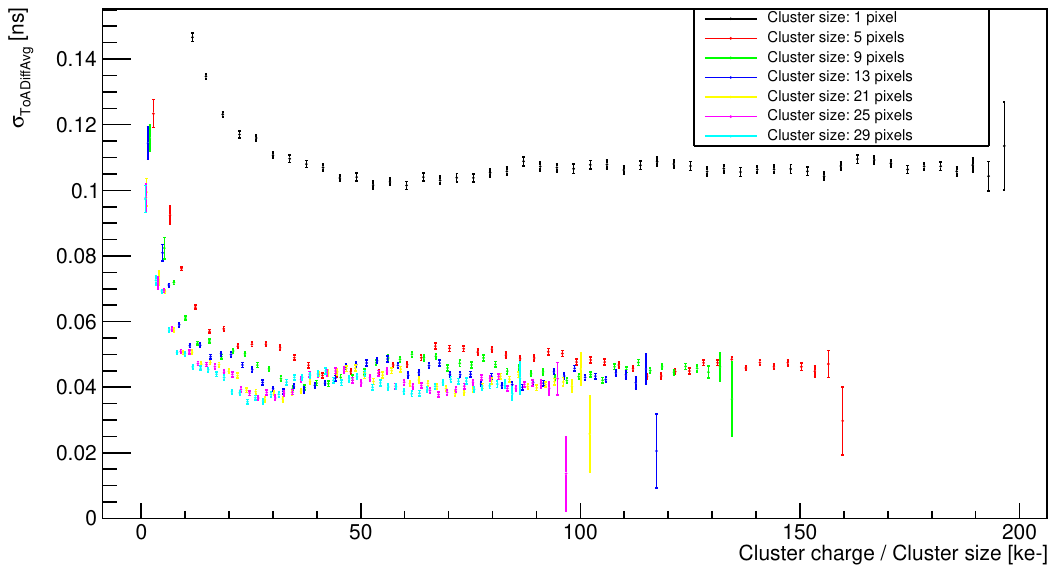}
\caption{Cluster timing resolution as a function of the cluster charge, divided by the cluster size. Different cluster sizes are presented as different colors. Timing resolution of the reference pixel has been subtracted. \label{fig:res_shell_dependence}}
\end{figure}

For each artificial cluster, the timing resolution distribution has been computed as a function of the cluster charge, as shown in Figure~\ref{fig:res_shell_dependence}. In order to better compare the results, cluster charge has been re-scaled dividing it by the cluster size.
The distribution corresponding to a single-pixel cluster is consistent with the one described in  Section $\ref{sec:Single_pixel_timing_res}$. 
Figure~\ref{fig:res_shell_dependence} shows two important contributions to the cluster timing resolution. 
The first is the dependence of the resolution on the cluster charge: increasing the total charge improves initially the timing resolution, as expected from the front-end jitter contribution, but for large charge values it leads to a worsening, explained in Section \ref{sec:Single_pixel_timing_res}.
The second contribution is due to the cluster size. A significant improvement in the timing resolution has been observed passing from a single-pixel cluster to a 5-pixel cluster, while the improvement given by a further increase in cluster size is modest. Thus, this proves that it's possible to reach the resolution obtained in the previous paragraph even with a lower multiplicity cluster.

\begin{figure}[htbp]
\centering
\includegraphics[width=.8\textwidth]{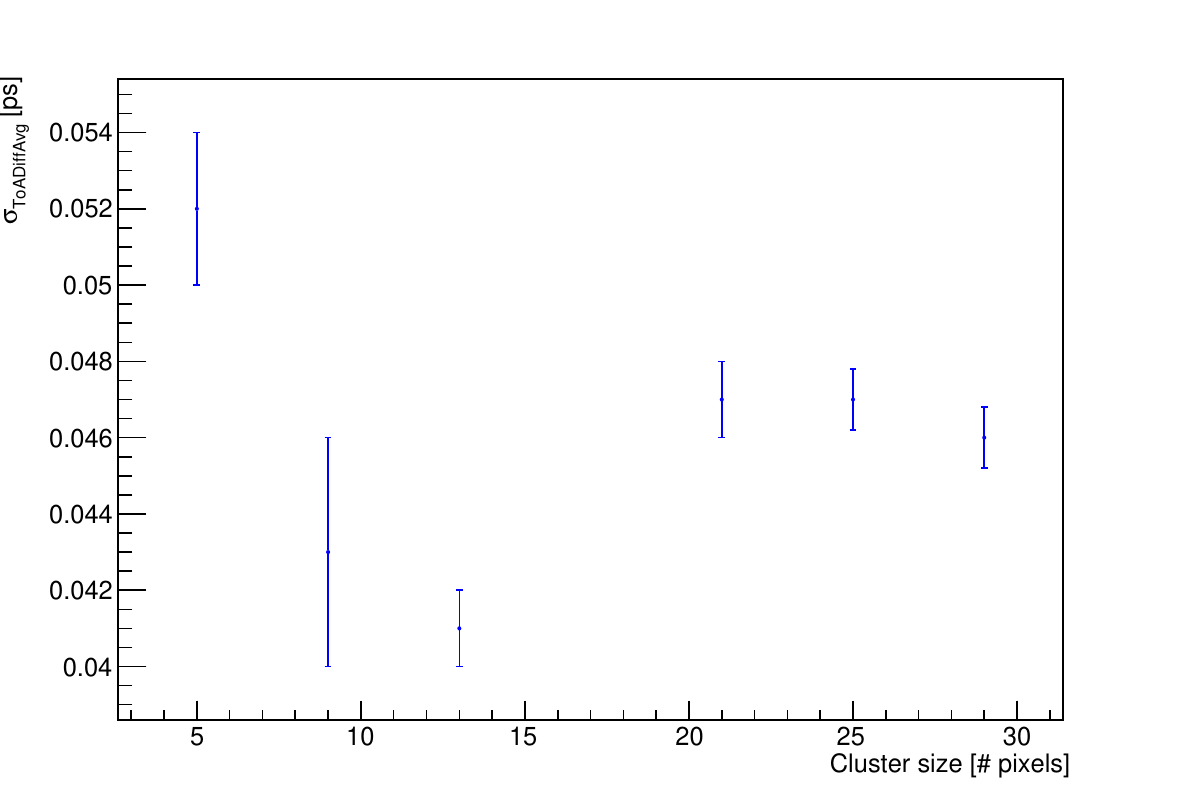}
\caption{Example of the cluster timing resolution dependence on the cluster size for a fixed range of cluster charge. A cut between $360\;\kilo\textrm{e}^-$ and $380\;\kilo\textrm{e}^-$ on the cluster charge has been placed.  Timing resolution of the reference pixel has been subtracted. \label{fig:shell_contribution_fixed_clusterToT}}
\end{figure}

In order to decouple the cluster charge and the cluster size contributions, clusters with different sizes but with ``fixed'' charge have been analysed, as shown for example in Figure \ref{fig:shell_contribution_fixed_clusterToT}, where only clusters with charge in the range $\left[360\;\kilo\textrm{e}^-;\;380\;\kilo\textrm{e}^-\right]$ have been included. This is equivalent to selecting a fixed total amount of charge spread over a different number of pixels.
Increasing the cluster size has two counteracting effects: first, the charge is spread among a greater number of pixels, thus each of them receives less charge, resulting in a worsening of the timing resolution due to the increased front-end jitter; second, the statistical factor due to the average improves the resolution of the cluster, as already demonstrated.

In Figure \ref{fig:shell_contribution_fixed_clusterToT} the first factor is dominant for larger cluster sizes, because the charge would spread more, and the average factor would not be able to compensate for the loss of resolution in each pixel due to the increased jitter. On the contrary, the improvement due to the average is dominant for low cluster sizes, because each pixel collects more charge and therefore the jitter contribution is much reduced.

Figure \ref{fig:shell_contribution_fixed_clusterToT} has been obtained from a "vertical" slice of Figure \ref{fig:res_shell_dependence} for a cluster charge included in the range $\left[360\;\kilo\textrm{e}^-;\;380\;\kilo\textrm{e}^-\right]$. 
Therefore, depending on the actual sensor coupled to the Timepix4 ASIC and its geometry, it will be possible to find the optimal cluster size in order to have the best timing resolution, depending on the expected total amount of charge that will be spread onto the Timepix4 matrix.

\begin{figure}[htbp]
\centering
\includegraphics[width=.9\textwidth]{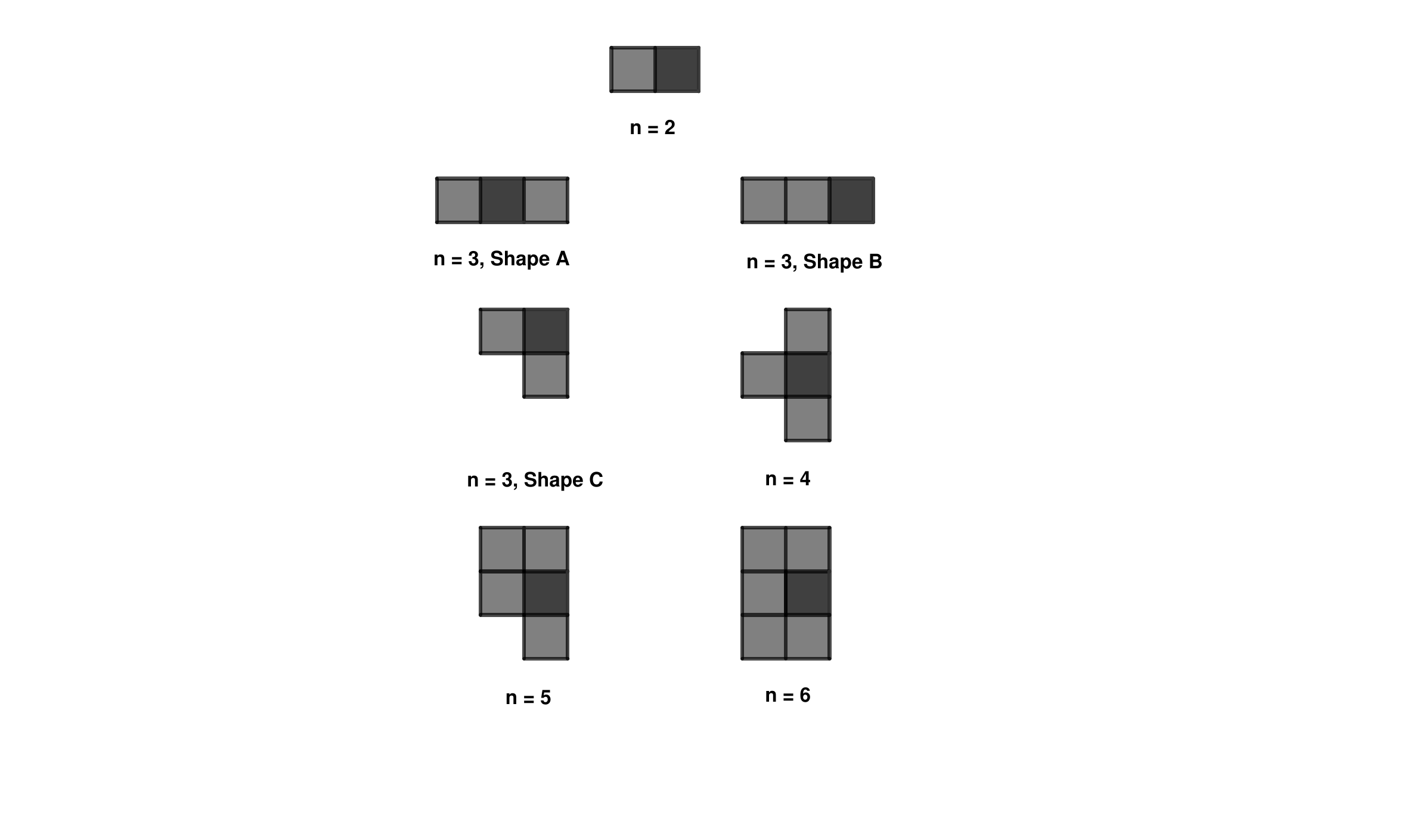}
\caption{Sketch showing clusters of different size (n) with asymmetric shapes artificially created via software. The darkest pixel on each cluster represents the one receiving the highest amount of charge within the cluster.\label{fig:shells_asymmetric}}
\end{figure}

Analogously, the analysis has been repeated for artificial clusters with asymmetric shapes, as shown in Figure \ref{fig:shells_asymmetric}, where the darkest pixel represents the one which received the highest amount of charge.

\begin{figure}[htbp]
\centering
\includegraphics[width=.7\textwidth]{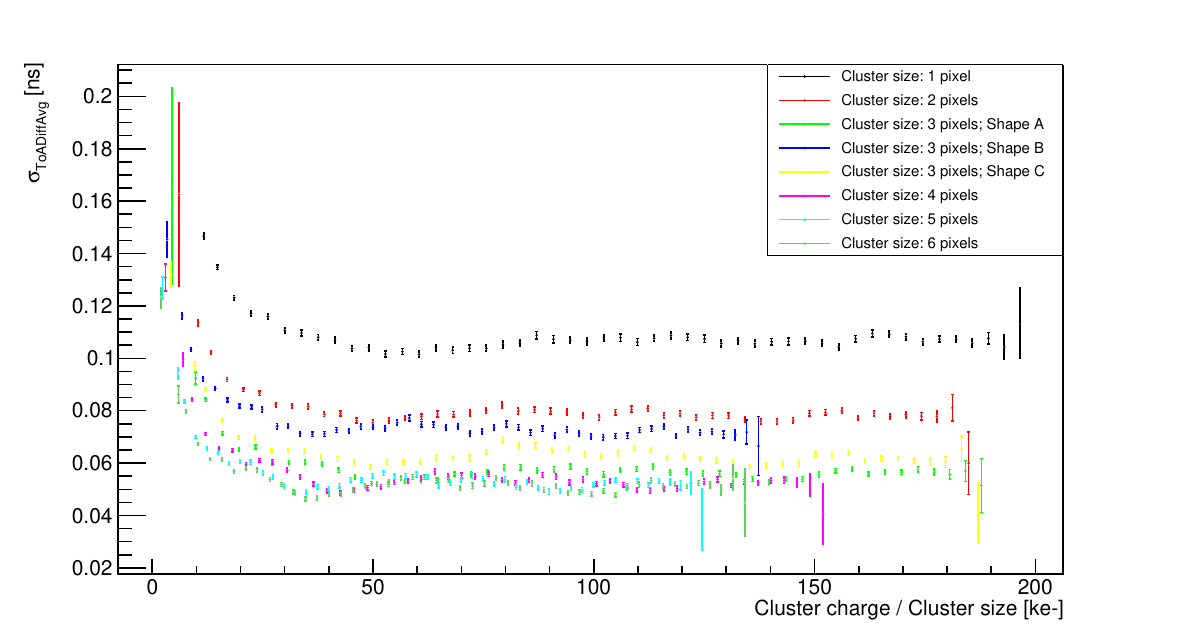}
\caption{Cluster timing resolution of clusters with asymmetric shapes as a function of the cluster charge, divided by the cluster size. Different cluster sizes are presented as different colors. Timing resolution of the reference pixel has been subtracted. \label{fig:res_shell_dependence_asym_clusters}}
\end{figure}

The resulting timing resolution distributions obtained once the reference resolution has been subtracted, presented in Figure \ref{fig:res_shell_dependence_asym_clusters}, show again an evident improvement with respect to the single pixel timing resolution, and again a modest one when increasing further the cluster multiplicity. In particular, it can be noticed that 3-pixel clusters with a "compact" shape (e.g. shapes A and C, where the pixel with the highest amount of charge is in a central position) present a resolution compatible with clusters of multiplicity 4, 5 or 6.

\section{Conclusions}\label{sec:Conclusions}

The timing performance of the Timepix4 ASIC has been extensively characterized using a picosecond laser setup illuminating a bump-bonded assembly based on $100\;\micro\meter$ n-on-p silicon sensor. 

Reaching the expected time performance on every pixel of the whole matrix requires careful calibration and correction procedures.
The Timepix4 timing resolution is heavily affected by the different VCO oscillation frequencies, and an automated calibration procedure had been implemented to correct for these variations on the whole pixel matrix. Another automated correction is implemented to correct for the time walk effect in  each individual pixel.

The single pixel timing resolution amounts to $\sigma_{\textrm{pixel}} = 107\pm3\;\pico\second\;\textrm{r.m.s.}$ for signals above $50\;\kilo\textrm{e}^-$. When considering clusters of sufficient size, the best resolution of  $\sigma_{\textrm{cluster}} = 33\pm3\;\pico\second\;\textrm{r.m.s.}$ has been obtained. 
For input charges $\textrm{Q}_{\textrm{in}}\gtrsim20\;\kilo\textrm{e}^-$ the timing resolution is almost constant, excluding the contribution due to the measurements overlap, proving that the system resolution is dominated by the TDC resolution, and no more by the front-end one.

We have demonstrated that cluster timing resolutions of few tens of picoseconds could be reached even when small size clusters are used, provided that the average charge received per pixel is $\gtrsim 20\;\kilo\textrm{e}^-$, therefore validating the concept of the 4DPHOTON detector.

\appendix

\acknowledgments

This work was carried out in the context of the Medipix4 Collaboration based at CERN, and in the framework of the MEDIPIX4 project funded by INFN CSN5.
This project has received funding from the European Research Council (ERC) under the European Union's Horizon 2020 research and innovation programme (Grant agreement No. 819627, 4DPHOTON project).

\bibliographystyle{JHEP}
\bibliography{biblio.bib}

\end{document}